\documentclass{amsart}

\usepackage{amsmath}
\usepackage{mathtools}
\usepackage{amssymb}
\usepackage{enumerate}
\usepackage{slashed}
\usepackage{graphicx}
\usepackage{newlfont}
\usepackage{amsrefs}
\usepackage{comment}
\usepackage[normalem]{ulem}
\usepackage{mathtools}
\usepackage{accents}
\usepackage{leftidx}

\numberwithin{equation}{section}

\theoremstyle{plain}
\newtheorem{theorem}{Theorem}
\newtheorem{lemma}{Lemma}

\newtheorem{cor}{Corollary}

\newtheorem{proposition}[theorem]{Proposition}

\theoremstyle{definition}

\theoremstyle{remark}
\newtheorem{remark}{Remark}

\renewcommand{\div}{\operatorname{div}}

\begin{document}

\title[The Penrose inequality and positive mass theorem with charge for cylindrical manifolds]{The Penrose inequality and positive mass theorem with charge for manifolds with asymptotically cylindrical ends}

\author[Jaracz]{Jaroslaw S. Jaracz}
\address{Department of Mathematics\\
Texas State University\\
San Marcos, TX 78666, USA}
\email{jsj74@txstate.edu}

\begin{abstract}
We establish the charged Penrose inequality 
\begin{equation}\nonumber
m\geq \frac{1}{2}\left( \rho + \frac{q^2}{\rho}   \right)
\end{equation}
for time symmetric initial data sets having an outermost minimal surface boundary and finitely many asymptotically cylindrical ends, with an appropriate rigidity statement. This is accomplished by a doubling argument based on the work of Weinstein and Yamada \cite{WY}, and a subsequent application of the ordinary charged Penrose inequality as established by Khuri, Weinstein, and Yamada \cite{KhuriWeinsteinYamada1, KhuriWeinsteinYamada2}. Furthermore, the techniques used in the aforementioned proof allow for an alternative proof of the positive mass theorem with charge for such manifolds, a result originally obtained in \cite{ChruscielBartnik}. 
\end{abstract}
\maketitle

\section{Introduction}
\label{sec1} \setcounter{equation}{0}
\setcounter{section}{1}

One of the most famous inequalities of mathematical general relativity is the Penrose inequality. Conjectured by Penrose \cite{Penrose} in the early 1970's using a heuristic argument based on the establishment viewpoint of gravitational collapse and the assumption of cosmic censorship, the Penrose inequality relates the mass $m$ to the surface area $A$ of the black hole. Defining the area radius $\rho$ by $A=4\pi \rho^2$ the Penrose inequality takes the form 
\begin{equation} \label{penrose}
m\geq \frac{1}{2}\rho.
\end{equation}

The Riemannian Penrose inequlaity is a special case, where the mass $m$ is the ADM mass of an asymptotically flat 3-manifold having non-negative scalar curvature and $A$ is the area of the outermost minimal surface (with possibly multiple components). It was proven in the late 1990's by Huisken and Ilmanen using weak Inverse Mean Curvature Flow (IMCF) with the area $A$ being the largest connected component of the outermost minimal surface \cite{HuiskenIlmanen}, and in full generality by Bray using a novel conformal flow of metrics \cite{Bray}. 

In the case that the outermost minimal surface has a single boundary component, inequality \eqref{penrose} can be extended to include charge yielding
\begin{equation} \label{charged penrose}
m\geq \frac{1}{2}\left( \rho + \frac{q^2}{\rho}    \right)
\end{equation}
where $q$ is the total charge enclosed by the apparent horizon. One might then conjecture that \eqref{charged penrose} holds when the outermost apparent horizon has multiple components. However, a time symmetric, asymptotically flat counter example was constructed in \cite{WY} by gluing two copies of the Majumdar-Papapetrou initial data sets. It is important to point out that this does not provide a counter example to cosmic censorship. As pointed out by Jang \cite{jang}, \eqref{charged penrose} is equivalent to two inequalities, 
\begin{equation} \label{penrose inequalities}
m-\sqrt{m^2-q^2}\leq \rho \leq m+\sqrt{m^2-q^2}
\end{equation}  
and only the upper bound follows from Penrose's heuristic arguments. The counter example violates the lower bound.

The usual geometry associated with the Penrose inequality is that of a manifold with boundary, where the boundary is a compact outermost apparent horizon (which coincides with the an outermost minimal surface in the time-symmetric setting). The typical examples of such geometry are the canonical slices of the Schwarzchild and non-extreme Reissner-Nordstr\"{o}m spacetimes.  One can consult any standard general relativity text such for the details. The case where $m=|q|$ is called extreme Reissner-Nordstr\"{o}m and is a complete manifold without boundary having the topology of $\mathbb{R}\times S^2$. Using a coordinate transformation, the metric can be put into the form  
\begin{equation}
\frac{(e^{-t/m}+m)^2}{m^2}dt^2+(e^{-t/m}+m)^2 d\sigma^2
\end{equation}
where $d\sigma^2$ is the metric on the Euclidean unit sphere. We see that letting $t\rightarrow \infty$, the metric approaches
\begin{equation}
dt^2+m^2d\sigma^2.
\end{equation}  

Thus, the region corresponding to large $t$ is referred to as an asymptotically cylindrical end, since in the limit the metric is a product of the ordinary metric on $\mathbb{R}$ and a metric on the cross section $S^2$. In addition, the mean curvature of the cross sections approaches $0$ as $t\rightarrow \infty$, and so in the limit one would expect the end to act like a compact outermost minimal surface. This is what motivates the subsequent definition of asymptotically cylindrical ends.

\subsection{General Initial Data Sets}
An initial data set for the Einstein-Maxwell equations consists of a quintuple $(M, g, k, E, B)$ where $M$ is a smooth 3-manifold, $g$ is a Riemannian metric, $k$ is a symmetric 2-tensor (the extrinsic curvature) and $E$ and $B$ are vector fields representing the electromagnetic field. 
This initial data set satisfies the constraint equations 
\begin{align} \label{ConstraintEquations}
\begin{split}
16\pi\mu=R+(Tr_gk)^2-|k|^2\\
8\pi J=\div (k-(Tr_g k)g)
\end{split}
\end{align}
where $\mu$ and $J$ are the energy and momentum densities of the matter fields. It is often useful to subtract off the contributions to these quantities arising from the electromagnetic field, and refer to these quantities as $\mu_{EM}$ and $J_{EM}$ which yields

\begin{align} \label{ChargedConstraint}
\begin{split}
16\pi\mu_{EM}=16\pi \mu-2(|E|^2+|B|^2)\\
8\pi J_{EM}=8\pi J+2(E\times B).
\end{split}
\end{align}

We will focus on the time symmetric case defined by the condition $k=0$, and subsequently write our initial data sets as $(M, g, E, B)$. The constraint equations greatly simplify in this case. We will assume that our data satisfies the charged dominant energy condition (CDEC)
\begin{equation} \label{CDEC}
R\geq 2\left(|E|^2+|B|^2   \right).
\end{equation}
This should be compared with the usual assumed energy conditions. Ordinarily, the data is said to satisfy the dominant energy condition if $\mu \geq |J|$ and the charged dominant energy condition if $\mu_{EM} \geq |J_{EM}|$. Thus, our definition of the charged dominant energy condition is slightly weaker than the ordinary definition, but it proves to be sufficient for our purposes. 

The importance of the time symmetric case partially lies in how apparent horizons are represented in such initial data sets. A black hole is an object which can only be detected by knowing the full long-term evolution of the initial data set, which is generally a difficult problem. So, we focus on questions which can be answered in terms of the initial data alone. Given an orientable surface $S\subset M$ in our initial data set, we can compute what are called the null expansions 
\[
\theta_{\pm}=H\pm Tr_S k
\]
where $H$ is the mean curvature of $S$ and $Tr_S k$ is the trace of $k$ restricted to $S$ with respect to the metric induced on $S$. An apparent horizon is a surface where $\theta_{\pm}=0$ and it conforms with the expected behavior at the surface of a black hole. As the above formula shows, in the time symmetric case apparent horizons coincide with minimal surfaces. See \cite{Mars} for an introduction.

\subsection{Strongly Asymptotically Flat Ends}    

A manifold is said to have a strongly asymptotically flat (SAF) end if there is a region $\Omega$ diffeomorphic to the complement of a ball in $\mathbb{R}^3$, and in the coordinates given by this diffeomorphism the following fall off conditions hold:
\[
|\partial^n(g_{ij}-\delta_{ij})|=O(|x|^{-n-1}),\quad |\partial^nk_{ij}|=O(|x|^{-n-2}),\quad |\partial^nE^i|=O(|x|^{-n-2}), 
\]
\[
|\partial^nB^i|=O(|x|^{-n-2}), \quad n=0, 1, 2 \quad \text{as} \quad |x|\rightarrow \infty.
\]
We will denote the image of $\Omega$ under the above diffeomorphism by $\tilde{\Omega}=\lbrace x: |x|>r_0\rbrace \subset \mathbb{R}^3$ for some $r_0>0$. 
We also require that the scalar curvature $R$ satisfies $R\in L^1(\Omega)$.
Given such an end, we can compute its ADM mass which is given by 
\begin{equation} \label{ADM}
m=\lim_{r\rightarrow \infty}\frac{1}{16\pi} \int_{S_r} (g_{ij,i}-g_{ii,j})\nu^j dS
\end{equation}
where $S_r$ are coordinate spheres of radius $r$ in $\tilde{\Omega}$ and $\nu^j$ is the outward unit normal \cite{adm}. It is well known that with the above fall-off conditions this quantity is a geometric invariant of the given end \cite{bartnik}. Also, the above fall off conditions guarantee that the electric and magnetic charges measured at infinity given by   
\begin{equation} \label{charges}
q_e=\lim_{r\rightarrow \infty}\frac{1}{4\pi} \int_{S_r} E_j\nu^j dS, \quad q_b=\lim_{r\rightarrow \infty}\frac{1}{4\pi} \int_{S_r} B_j\nu^j dS 
\end{equation}
are well defined. We also define the squared total charge by
\begin{equation} \label{SquaredTotalCharge}
q^2=q_e^2+q_b^2.
\end{equation}

Given a SAF end we have the intuitive notion of one surface enclosing another. We can make this precise as follows. Conformally compactify the chosen SAF end by adding $\infty$, a point at infinity. Then given two smooth compact surfaces $S_1$ and $S_2$, we say that $S_1$ encloses $S_2$ if for any smooth curve $\gamma$ in the conformal compactification passing through $\infty$ and intersecting the image of $S_2$, $\gamma$ also intersects the image of $S_1$. 

\subsection{Asymptotically Cylindrical Ends}
Motivated by extreme Reissner-Nordstr\"{o}m, we define asymptotically cylindrical (AC) ends as follows. A region $\Lambda$ in our initial data set is said to be asymptotically cylindrical if it is diffeomorphic to $[0, \infty) \times S^2$ and there exists a product metric $\tilde{g}=dt^2+h$, with $h$ a Riemannian metric on $S^2$, such that in the coordinates given by the diffeomorphism
\[
|\partial^n(g_{ij}-\tilde{g}_{ij})|=O\left(1/t\right), \quad n=0, 1, 2 \quad t\rightarrow \infty.
\]
We will denote the image of $\Lambda$ under the diffeomorphism by $\tilde{\Lambda}$ and the resulting local coordinates by $(t, \omega)$. In addition, we require that if there are vector fields $E$ and $B$ defined on our cylindrical end, then there exist vector fields $\tilde{E}$ and $\tilde{B}$, with components independent of $t$, satisfying \[|\partial^n (E^i-\tilde{E}^i)|=O(1/t), \quad |\partial^n (B^i-\tilde{B}^i)|=O(1/t), \; \text{for} \; n=0,1 \; \text{as} \; t\rightarrow \infty. 
\]

We refer to the surfaces $N_T=\lbrace (t, \omega): t=T\rbrace$ as cross sections. With the above asymptotics, the areas of the cross sections approach a limiting value $A$ which for brevity we will refer to as the area of the AC end. As can be easily seen, this area is independent of the choice of AC coordinates. In addition, the mean curvatures of the cross sections tend to $0$ as $t\rightarrow \infty$.  As $t$ increases, we say that we are moving down the AC end. We also define sets $\tilde{\Lambda}_s=\lbrace (t, \omega): t>s \rbrace$ which are diffeomorphic to $\Lambda_s \subset M$.

 We can also define the intuitive notion of what it means for a surface to enclose a cylindrical end. Given a SAF end, we can again add a point at $\infty$. We now say a compact smooth surface $S_1$ encloses the cylindrical end if any smooth curve $\gamma$ through $\infty$ and having unbounded $t$-coordinate in the AC coordinates intersects $S_1$.
 
 When we have multiple cylindrical ends indexed by $i$ with $1\leq i \leq n$ we will denote the relevant sets by $\Lambda^i$, $\Lambda_s^i$, etc.
 
\subsection{IMCF and the Charged Hawking Mass}

In addition, in order to prove our rigidity statement, we will employ weak IMCF. In particular, we use the level set formulation which seeks to find a solution of 
\begin{equation} \label{IMCF}
\text{div}_{M} \left(\frac{\nabla u}{|\nabla u|}   \right)= |\nabla u|
\end{equation} 
where $u:M \rightarrow \mathbb{R}$ is a real valued function.
The Hawking mass can be generalized to the charged Hawking mass which for a surface $S$ is given by
\begin{equation} \label{CHM}
M_{CH}(S)=\sqrt{\frac{|S|}{16\pi}}\left( 1 + \frac{4\pi q^2}{|S|} - \frac{1}{16\pi} \int_{S} H^2  \right)
\end{equation} 
where $|S|$ is the area of the surface and $q^2$ is as above. The ordinary Hawking Mass, denoted $M_H(S)$ can be recovered by setting $q^2=0$. We will also use the well known monotonicity of the Hawking and charged Hawking masses under weak IMCF. For details, the reader is encouraged to consult \cite{jang}, \cite{HuiskenIlmanen}, and \cite{DisconziKhuri}.
 
\subsection{Statement of Results}

Our initial data set $(M, g, E, B)$ will consist of a manifold with a single SAF end, a boundary consisting of a compact minimal surface (with possibly multiple components), and finitely many AC ends. We will refer to the union of the ordinary boundary and the finitely many AC ends as the \textit{generalized boundary}. Denoting by $A_i$ the area of the $i$-th AC end, we will define
\begin{equation}
A=|\partial M| + \sum_{i=1}^{n} A_i
\end{equation}
to be the \textit{area of the generalized boundary}. We also define the \textit{area radius} $\rho$ of the generalized boundary by $A=4\pi \rho^2$. We will also say that a surface $S$ \textit{encloses the generalized boundary} if it encloses the ordinary boundary and each of the AC ends (in the appropriate sense described above). 

We also say that our data satisfies the \textit{Maxwell constraints without charged matter} if $\div E = \div B =0 $ everywhere on $M$.

In the statement of the Penrose inequality, it is assumed that the boundary is an outermost minimal surface because of the well known shielding effect where a large minimal surface can be hidden inside of a minimal surface of smaller area. The area of this large surface can then violate the Penrose inequality. See \cite{gibbons} for details. Noting that an outermost minimal surface is outerminimizing (which means that any surface enclosing it has strictly larger area), we make an analogous assumption that any surface $S$ enclosing the generalized boundary satisfies $|S|>A$. 

Khuri, Weinstein, and Yamada proved that given a strongly asymptotically flat initial data set $\left(M, g, E, B   \right)$ with outermost minimal surface boundary of area $A=4\pi \rho^2$, satisfying the charged dominant energy condition and maxwell constraints without charged matter, and with $|q|\leq \rho$, \eqref{charged penrose} holds, with equality if and only if the data arises as the canonical slice of the Reissner-Nordstr\"{o}m spacetime. Furthermore, when the condition $|q|\leq \rho$ is dropped, they proved that the upper bound in \eqref{penrose inequalities} holds, with equality if and only if the data is again the canonical slice of Reissner-Nordstr\"{o}m \cite{KhuriWeinsteinYamada2}. The goal of this paper is to extend their results to data sets containing AC ends, and their theorem will be the primary tool of doing so.

Our main result is the following:

\begin{theorem} \label{thm}

Let $(M, g, E, B)$ be a time symmetric initial data set with a single SAF end of ADM mass $m$, with generalized boundary consisting of finitely many AC ends and a minimal surface boundary, with generalized area radius $\rho$. Assume the data satisfies the charged dominant energy condition, the Maxwell constraints without charged matter, and has electric and magnetic charges $q_e$ and $q_b$ (with total squared charge $q^2$). Finally, suppose that the generalized boundary is outerminimizing, which means that any surface $S$ enclosing the generalized boundary satisfies $|S|> 4\pi \rho^2$. Then the upper bound in (\ref{penrose inequalities}) holds. Moreover, if $\rho \geq |q|$ then (\ref{charged penrose}) holds. Equality holds in both cases if and only if the initial data is given by the canonical slice of the (possibly extreme) Reissner-Nordstr\"om spacetime with $E=(q_e/r^2) \nu_r$ and $ B=(q_b/r^2)\nu_r$ where $\nu_r$ is the outward unit normal to spheres of radius $r$ in standard coordinates.

\end{theorem}

The techniques used in proving the above theorem also allow us to extend the positive mass theorem with charge to initial data sets containing AC ends. 

\begin{theorem} \label{thm2}
	Let $(M, g, E, B)$ be a time symmetric initial data set for the Einstein-Maxwell equations with a single SAF end with a generalized boundary consisting of finitely many asymptotically cylindrical ends and a minimal surface boundary. Then
	\begin{equation}
	m\geq |q|
	\end{equation}
	where $m$ is the ADM mass and $q^2$ the squared total charge of the SAF end. 
\end{theorem}

\section{Proof of Theorem 1}

\subsection{Sketch of the Proof}

The main idea of the proof is to use Theorem 1.1 and Corollary 1.2 in \cite{KhuriWeinsteinYamada2} due to Khuri, Weinstein, and Yamada. For brevity we will refer to Theorem 1.1 and Corollary 1.2 as the KWY theorem. However, our initial data set currently does not satisfy all the hypotheses of this theorem. Specifically, it does not have a minimal surface boundary with area $A$. 

We remedy this by doubling across the AC ends. Specifically, we go sufficiently far down the ends and take what we call a collar neighborhood in each end. We then chop off the ends beyond these collars, and obtain transitions to the corresponding cylindrical metrics in each collar, giving us a manifold which we denote by $M^+$. So, for each cylindrical end we have a new boundary component labeled by $S^{+}_{i}$. Now, we take a second copy of $M^+$ denoted $M^-$. We identify the boundary components $S^{+}_{i}$ and $S^{-}_{i}$ giving us a smooth Riemannian manifold since the metric near each of these components is the appropriate cylindrical metric. We denote this smooth Riemannian manifold by $\hat{M}$, and its metric by $\hat{g}$.  Outside of the collars, the metric is the same as our original metric, and so $\hat{M}$ is a manifold with two SAF ends. The image under the gluing of the $S^{\pm}_i $ is now the fixed point set of an isometry with respect to $\hat{g}$ metric, and is therefore a minimal surface. 

We can't directly push forward our electromagnetic fields to $\hat{M}$ since the resulting vector field won't be continuous. However, it is also obvious we don't need them to be defined on the entire manifold. We push them forward only to $\hat{M}^+$ (the image of $M^+$ under the gluing), so that they are defined exterior to the outermost minimal surface with respect to that end. Unfortunately, these vector fields are no longer divergence free in the transition region, and so no longer satisfy the assumptions of the KWY theorem. Hence, we perturb the electromagnetic fields to restore the divergence constraint. This perturbation changes the charge, so we then check that the further down the AC ends we perform the chopping, the closer this new charge is to the old one.

In addition, our initial data no longer satisfies the CDEC. We restore it by a conformal perturbation of $\hat{g}$ where the conformal factor $u$ satisfies an appropriate elliptic equation. We check that our initial data still possesses a minimal surface, and apply the KWY theorem to the resulting data. We then check that the ADM mass and charge of this conformal data can be made arbitrarily close to the original values by chopping further and further down the AC ends, and take the limit to obtain the desired result. 

Finally, we establish the rigidity result by employing weak IMCF. By the arguments of \cite{KhuriWeinsteinYamada2} if equality holds the generalized boundary must consist of either an $S^2$ outermost minimal surface (in which case the data is Reissner-Nordstr\"om) or a single AC end. In the latter case, we prove that there exists a smooth IMCF which provides a diffeomorphism between our original data and the canonical slice of the extreme Reissner-Nordstr\"om spacetime.     

For technical purposes, we begin by showing that we can conformally deform our data to satisfy the strict CDEC, with mass, charge, and generalized boundary area arbitrarily close to their original values. We then prove the charged Penrose inequality for this deformed data and then let the mass, charge, and area approach their original values, establishing the charged Penrose inequality for the original data.

\subsection{Definition of Weighter H\"{o}lder Spaces} \label{DefinitionHolderSpace} First we will define weighted H\"{o}lder spaces on a manifold with finitely many SAF ends. Denote each of the SAF ends by $\Omega_k$. For each end, fix a choice of SAF coordinates. We write $\tilde{\Omega}_k=\lbrace x: |x|>r_{0, k} \rbrace$. We also define $\tilde{\Omega}_k(s)=\lbrace x: |x|>s>r_{0, k} \rbrace \subset \tilde{\Omega}_k$ and the images of these sets under the inverse diffeomorphism by $\Omega_k(s)$.  Let $K=M\setminus \left( \cup_k \Omega_k(r_{0, k}+2)\right)$. Define a smooth function $\sigma \geq 1$ by $\sigma \equiv 1$ on $K$, and such that sufficiently far in each $\Omega_k$ we have $\sigma=r$, the Euclidean radial coordinate in each of the chosen SAF coordinates. This $\sigma$ is called the weight function. Notice that if $M$ has AC ends, then $\sigma=1$ (sufficiently far down) in these ends. 

We would like to define the weighted H\"{o}lder spaces in the same way as was done in \cite{WY}, by defining them to be the set of functions $\phi$ on $M$ whose $k$-th order derivatives are H\"{o}lder continuous and for which the norm $\Vert \phi \Vert_{C^{k, \alpha}_{-\beta}}$ defined below is finite:
\[
\Vert \phi \Vert_{C^{k}_{-\beta}}=\sum_{i=0}^{k} \left\Vert \sigma^{\beta+i} D^i \phi \right\Vert_{C^0} 
\]

\[ \left[ D^k \phi  \right]_{\alpha, -\beta}= \sup_{0<\text{dist}(x, y)<\rho} \sigma(x, y)^{k+\beta+\alpha} \frac{|P^x_y D^k \phi(y)-D^k \phi(x)|}{\text{dist}(x, y)^\alpha}
\]
\[
\Vert \phi \Vert_{C^{k, \alpha}_{-\beta}}=\Vert \phi \Vert_{C^{k}_{-\beta}}+\left[ D^k \phi  \right]_{\alpha, -\beta}.
\]
Here $D^i \phi$ is the tensor of $i$-th order derivatives of $\phi$, $\rho$ is the injectivity radius of $M$, $\sigma(x, y)=\max{\sigma(x), \sigma(y)}$, and $P^x_y$ is parallel translation along the shortest geodesic from $y$ to $x$. 

This is a nice definition since it is coordinate independent. However, it does not quite make sense when the manifold in question has a boundary since then the injectivity radius is $0$. We remedy this by giving an alternative definition as follows. 

Suppose $M$ has $m$ SAF ends and $n$ AC ends. The cross section of the $i$-th AC end is $S^2$ and hence can be covered by two sets diffeomorphic to open disks. Hence using local coordinates we can cover each $\Lambda^i_3$ by sets $V_{i, 1}$ and $V_{i, 2}$ which are diffeomorphic to $\tilde{V}_{i, k}=D\times (2, \infty)\subset \mathbb{R}^3$ where $D$ is the standard (open) unit disk. 

Next, we consider the compact set $K=M\setminus \left( \Omega_1 \cup \dots \cup \Omega_m \cup \Lambda_3^1 \cup \dots \cup \Lambda_3^n  \right)   $. We can cover $K$ by finitely many pre-compact open sets diffeomorphic to standard balls and half balls (near $\partial M$) in $\mathbb{R}^3$. We can also choose the coordinates in such a way that $\partial_{x^3}|_{x^3=0}=\partial_{\tau}$, the normal derivative with respect to the Riemannian metric induced on the half-ball from $(M, g)$. We denote these sets by $W_i$ with $1\leq i \leq w$ and their diffeomorphic images as $\tilde{W}_i \subset \mathbb{R}^3$.  

The collection of sets $\lbrace \Omega_1, \dots \Omega_m, V_{1, 1}, V_{1, 2}, \dots V_{n,1}, V_{n, 2}, W_1, \dots W_w \rbrace$ forms a finite open cover of $M$ by sets diffeomorphic to open subsets of $\mathbb{R}^3$. Given a function $\phi$ defined on $M$ we can then pull it back to each of theses open subsets. We can also pull back $\sigma$. By a slight abuse of notation we continue to denote these pulled back functions by $\phi$ and $\sigma$. 

For the $i$-th SAF end we define 
\[
\Vert \phi \Vert_{C^{k}_{-\beta}(\tilde{\Omega}_i)}=\sum_{i=0}^{k} \sup_{|\gamma|=i} \sup_{x\in \tilde{\Omega}_i} |\sigma^{\beta+i}D^\gamma \phi|  
\]

\[ \left[ D^k \phi  \right]_{\alpha, -\beta}^i= \sup_{|\gamma|=k} \sup_{x, y\in \tilde{\Omega}_i  }  \sigma(x, y)^{k+\beta+\alpha} \frac{|D^\gamma \phi(y)-D^\gamma \phi(x)|}{|x-y|^{\alpha}}
\]
\[
\Vert \phi \Vert_{C^{k, \alpha}_{-\beta}(\tilde{\Omega}_i)   }=\Vert \phi \Vert_{C^{k}_{-\beta}(\tilde{\Omega}_i)}+\left[ D^k \phi  \right]_{\alpha, -\beta}^i.
\]  
In addition, for a smooth open domain of $U \subset \mathbb{R}^3$ we can compute the ordinary H\"{o}lder norm $\Vert \cdot \Vert_{C^{k, \alpha}(U)} $ (see \cite{GT}). Therefore we define 

\begin{equation} \nonumber
\Vert \phi \Vert_{C^{k, \alpha}_{-\beta}(M)} = \sum_{i=1}^m \Vert \phi \Vert_{C^{k, \alpha}_{-\beta}(\tilde{\Omega}_i)   } + \sum_{i=1}^n \left( \Vert \phi \Vert_{C^{k, \alpha}(\tilde{V}_{i, 1})   } + \Vert \phi \Vert_{C^{k, \alpha}(\tilde{V}_{i, 2})   }   \right) + \sum_{i=1}^w \Vert \phi \Vert_{C^{k, \alpha}(\tilde{W}_i)   }. 
\end{equation} 
To summarize, we take a finite open cover of $M$, compute the appropriate weighted or unweighted H\"{o}lder norms in local coordinates, and take the sum. We define $C^{k, \alpha}_{-\beta}$ to be the class of functions for which the above defined norm is finite. Even though the norm depends on the choice of open cover, it can be checked that the class of functions itself does not.

We will also need a different class of functions with appropriate weights in the AC ends. As before, fixing a choice of AC coordinates on each of our AC ends we define a set $K_2=M\setminus \cup_i \Lambda_1^i$. Now let $\mu\geq 1$ be a smooth function such that $\mu \equiv 1$ in $K_2$ and such that sufficiently far down each AC end $\mu=e^{\delta t}$, the local coordinate in $\tilde{\Lambda}^i$. Now let $C^{k, \alpha}_{-\beta, \delta}$ be the set of functions $\phi$ on $M$ whose $k$-th order derivatives are H\"{o}lder continuous and for which the norm $\Vert \phi \Vert_{C^{k, \alpha}_{-\beta, \delta}}$ defined below is finite:  
\[
\Vert \phi \Vert_{C^{k, \alpha}_{-\beta, \delta}}=\Vert \mu \phi \Vert_{C^{k, \alpha}_{-\beta}}.
\]

As usual, the spaces we have defined so far are Banach spaces. We also define certain important closed subsets of these. Define
\[      
\tilde{C}^{k, \alpha}_{-\beta}=\left\lbrace \phi\in C^{k, \alpha}_{-\beta}: \partial_{\tau} \phi|_{\partial M}=0   \right\rbrace
\]
and similarly
\[      
\tilde{C}^{k, \alpha}_{-\beta, \delta}=\left\lbrace \phi\in C^{k, \alpha}_{-\beta, \delta}: \partial_{\tau} \phi|_{\partial M}=0   \right\rbrace
\] 
where $\partial_{\tau}$ denotes the normal derivative. Being closed subsets of Banach spaces, these are Banach spaces themselves. 

\subsection{A Conformal Deformation of the Initial Data} \label{InitialConformalDeformation} 

In order to prove some subsequent technical propositions, we will actually need our initial data set to satisfy the strict charged dominant energy condition, $R>2(|E|^2+|B|^2)$. In order to do so, we will conformally deform our initial data set to one with mass, charge, and generalized boundary area $\varepsilon$-close to our original data. 

Suppose we have our data set $(M, g, E, B)$ and consider the conformal metric $\bar{g}=u^4 g$. We have the well known formula for the scalar curvature under a conformal change 
\begin{equation}\label{ScalarUnderConformalChange}
\bar{R}=u^{-4}(R-u^{-1}\Delta_g u) 
\end{equation} 
where $\Delta_g$ denotes the Laplace-Beltrami operator with respect to the $g$ metric. This suggests we solve the elliptic problem
\begin{align} 
\begin{split} \label{EP1}
\Delta_g u -\frac{1}{8}Ru+\frac{1}{8}\left(R+\varepsilon\varrho\right)u^{-3}=0 \quad \text{on}\quad M \\
\partial_{\tau}u=0 \quad \text{on} \quad \partial M, \quad u\rightarrow 1 \quad \text{as} \quad r \rightarrow \infty.
\end{split}
\end{align} 

By $r \rightarrow \infty$ we mean as the points move into both the SAF and AC ends. The function $\varrho$ is some smooth non-negative function which vanishes sufficiently fast as we move into the SAF and AC ends, and $\varepsilon>0$ is some small real parameter. It is easy to check that using such a $u$ as the conformal factor the scalar curvature is $\bar{R}=\left(R+\varepsilon\varrho\right)u^{-8}$. 

We define vector fields by $\bar{E}=u^{-6}E$ and $\bar{B}=u^{-6}B$. If $E$ and $B$ are divergence free, then so are $\bar{E}$ and $\bar{B}$ as shown by the following computation:
\begin{equation} \label{div}
\div_{\bar{g}}(\bar{E})=\frac{1}{\sqrt{|\bar{g}|}}\partial_i(\sqrt{|\bar{g}|}\bar{E}^i)=\frac{u^{-6}}{\sqrt{|g|}}\partial_i(\sqrt{|g|}E^i)=u^{-6}\div_g(E)=0
\end{equation} 
and similarly for $\bar{B}$. Furthermore
\begin{equation} \label{ScalarDeformation}
\bar{R}=(R+\varepsilon \varrho)u^{-8}\geq 2u^{-8}(|E|^2_g+|B|^2_g)+u^{-8}\varepsilon\varrho > 2(|\bar{E}|^2_{\bar{g}}+|\bar{B}|^2_{\bar{g}})
\end{equation}
and so this deformed data satisfies the strict CDEC. What remains to be verified is that we can find such a conformal factor $u$ which is close to $1$ in an appropriate sense, and so that the resulting quantities $\bar{m}$, $\bar{q}$, and $\bar{A}$ are close to the originals. 

Since we are looking for a small deformation of $1$ we will write $u=1+\phi$ for some small function. Then the above elliptic problem becomes 

\begin{align} 
\begin{split} \label{ellipticproblem} 
\Delta_g \phi -\frac{1}{8}R(1+\phi)+\frac{R}{8(1+\phi)^3}+\frac{\varepsilon\varrho}{8(1+\phi)^3}=0 \quad \text{on}\quad M \\
\partial_{\tau}\phi=0 \quad \text{on} \quad \partial M, \quad \phi\rightarrow 0 \quad \text{as} \quad r \rightarrow \infty
\end{split}
\end{align} 
and after a bit of rearrangement we can write \ref{ellipticproblem} as
\begin{equation} \label{ellipticproblem3} 
\Delta_g \phi -\frac{R}{8}\left( \frac{4\phi+6\phi^2+4\phi^3+\phi^4}{(1+\phi)}  \right) + \frac{\varepsilon\varrho}{8(1+\phi)^3}=0. 
\end{equation}
We will find a solution in the appropriate H\"{older} space by using the implicit function theorem for Banach spaces. 

\begin{proposition} \label{Isomorphism}
The operator
\begin{equation}
L=\Delta_g - \frac{1}{2}R \; : \; \tilde{C}^{2, \alpha}_{-2/3, \delta} \rightarrow C^{0, \alpha}_{-8/3, \delta}
\end{equation}
is an isomorphism for all $\delta \in \mathbb{R}_+ \setminus D$ where $D$ is some discrete set.
\end{proposition} 

\noindent \textbf{Proof:} Integrating by parts we see that $L$ is injective. As we go down each cylindrical end, $L$ approaches the translation invariant operator $L_0=D_t^2+\Delta_N+R_0$ where $\Delta_N$ is the Laplacian on the cross section with respect to the metric $h$ and $R_0$ is a function independent of $t$. Combining theorem 2.3.12 in \cite{Nordstrom} with the results of \cite{Choquet} the result follows.   $\qed$

Next consider defining an operator 
\begin{equation}
\mathcal{F}(\varepsilon, \phi)=\Delta_g \phi -\frac{1}{8}R(1+\phi)+\frac{R}{8(1+\phi)^3}+\frac{\varepsilon\varrho}{8(1+\phi)^3}
\end{equation}
with $\mathcal{F}: \mathbb{R} \times V \rightarrow C^{0, \alpha}_{-8/3, \delta}$ where $V=\lbrace \phi\in \tilde{C}^{2, \alpha}_{-2/3, \delta}: \Vert \phi \Vert_{\tilde{C}^{2, \alpha}_{-2/3, \delta}}<1/2   \rbrace$.
The fact that the above operator with the given domain has the given range can be seen by looking at it in the form \eqref{ellipticproblem3}. 
We can now apply the implicit function theorem.

\begin{proposition} \label{SolutionOfEllipticProblem}
For every $\epsilon>0$ there is an $\varepsilon>0$ such that the elliptic problem \eqref{ellipticproblem} has a solution $\phi$ satisfying $ \Vert \phi \Vert_{\tilde{C}^{2, \alpha}_{-2/3, \delta}}<\epsilon$.
\end{proposition}  

\noindent \textbf{Proof:} Consider $\mathbb{R} \times V$ and the operator $\mathcal{F}: \mathbb{R} \times V \rightarrow C^{2, \alpha}_{-8/3, \delta} $. Notice that $\mathcal{F}(0, 0)=0$ and the linearization of the operator is $d\mathcal{F}(0, 0)=\Delta_g - \frac{1}{2}R=L$ is an isomorphism by Proposition \ref{Isomorphism}. The claim now follows by the implicit function theorem. $\qed$

\medskip

The appropriate elliptic regularity of the solution can be obtained using Proposition 12B.1 in \cite{Taylor}. In fact, if we start with a smooth metric and choose a smooth function $\varrho$, then the resulting solution $\phi$ will be smooth for all sufficiently small $\varepsilon$.

We will also need the following lemma.

\begin{lemma} \label{ADMestimate1}
Suppose that $\phi\in C^{0, \alpha}_{-2/3}(\Omega)$ and $\Delta_{g}\in C^{0, \alpha}_{-3}(\Omega)\cap L^1(\Omega)$. Then there is a constant $C$ such that 
\begin{equation}
\Vert \phi \Vert_{C^{2, \alpha}_{-1}} \leq C \left( \Vert \Delta_{g} \phi\Vert_{C^{0, \alpha}_{-3}\cap L^1} + \Vert \phi \Vert_{C^{2, \alpha}_{-2/3}}  \right).
\end{equation}
\end{lemma}

\noindent \textbf{Proof:} The proof is given in Lemma 6 of \cite{WY}. $\qed$

\medskip

We can now prove the following:

\begin{proposition} \label{ConformalDeformation1}
Let $(M, g, E, B)$ be an initial data set satisfying the charged dominant energy condition and the Maxwell constraints without charged matter, with outermost minimal surface boundary and finitely many AC ends, having mass $m$, charge $q^2$ and area $A$. Then for any $\epsilon>0$ exists a conformal deformation such that the data set satisfies the same hypotheses, the strict charged dominant energy condition with mass $\bar{m}$, total charge $\bar{q}^2$ and area $\bar{A}$ such that $q^2=\bar{q}^2$, and there exists a constant $C>0$ such that $|\bar{m}-m|, |\bar{A}-A|<C\epsilon$. 
\end{proposition}

\noindent \textbf{Proof:} We consider the solution given by Proposition \ref{SolutionOfEllipticProblem}. Define $u=1+\phi$, and $\bar{g}=u^4g$. We define the vector fields $\bar{E}=u^{-6}$ and $\bar{B}=u^{-6}B$. As discussed earlier, these vector fields are divergence free and the resulting data set satisfies the strict charged dominant energy condition. Since $\partial_{\tau} u = \partial_{\tau} \phi =0$ the boundary remains minimal.

Let $S$ be an embedded hypersurface in $M$. If $dS_g$ is the induced volume form in the $g$ metric, then the induced volume form in $\bar{g}$ is given by $dS_{\bar{g}}=u^4 dS_g$. Since in the AC ends $\phi \sim e^{-\delta t}$ we see these ends remain AC with respect to $\bar{g}$ with the same area. Since $|\phi|<\epsilon$ we have 
\begin{equation} \nonumber
(1-5\epsilon)|\partial M|_g \leq (1-\epsilon)^4|\partial M|_g \leq |\partial M|_{\bar{g}} \leq (1+\epsilon)^4|\partial M|_g\leq (1+5\epsilon)|\partial M|_g
\end{equation} for sufficiently small $\epsilon$, and so $|\bar{A}-A|\leq 2|\partial M|_g \epsilon$. 

Noticing that if $\nu$ is the unit normal to a surface then $\bar{\nu}=u^{-2}\nu$, and using equations \eqref{charges} we compute
\begin{equation}
\bar{q}_e=\lim_{r\rightarrow \infty}\frac{1}{4\pi} \int_{S_r} \bar{E}_j \bar{\nu}^j dS_{\bar{g}} = \lim_{r\rightarrow \infty}\frac{1}{4\pi} \int_{S_r} (u^{-2})E_j (u^{-2})\nu^j (u^4)dS =\lim_{r\rightarrow \infty}\frac{1}{4\pi} \int_{S_r} E_j \nu^j dS = q_e
\end{equation}  
and similarly for $\bar{q}_b$. Hence, the charges are conserved. 

Finally, notice that 
\begin{equation}
\Delta_g \phi = \frac{R}{8}\left( \frac{4\phi+6\phi^2+4\phi^3+\phi^4}{(1+\phi)}  \right) - \frac{\varepsilon\varrho}{8(1+\phi)^3} \in C^{0, \alpha}_{-3} \cap L^1
\end{equation} 
in $\Omega$, and so by Lemma \ref{ADMestimate1} we have $\phi \in C^{2, \alpha}_{-1}$. Using the formula \eqref{ADM} it is then easy to show $|\bar{m}-m| \leq C\epsilon$ for some constant. $\qed$

\subsection{The Gluing}
\label{Gluing}

We index our $n$ cylindrical ends by $i$ with $1\leq i \leq n$. Each end has the topology of $[0, \infty) \times S^2$. On each particular AC end we fix a choice of cylindrical coordinates.

Start with the $i=1$ AC end. Denote by $\tilde{g}_1$ the product metric on $\mathbb{R} \times S^2$ which is the limit of $g$. Now, using these cylindrical coordinates we consider $M_1=M\setminus \lbrace (t_1, \omega): t_1>T \rbrace$; that is, our manifold with the cylindrical end chopped off beyond $t_1=T$ in the fixed cylindrical coordinates. We also consider the region $\Sigma_1(T)=\lbrace (t_1, \omega): T-3\leq t_1 \leq T  \rbrace$. 

Now, we consider a smooth cut-off function $\chi_1(t_1)$ defined in $\Sigma_1(T)$ such that $\chi_1 \equiv 1 $ for $T-3 \leq  t_1 \leq T-2$, decreasing for $T-2< t_1 < T-1$ and $\chi_1 \equiv 0 $ for $T-1 \leq  t_1 \leq T$. Next, we define $\chi_2(t_1)=1-\chi_1(t_1)$. Finally, we define a new metric $g_{1}$ on $\Sigma_1(T)$ by $g_{1}=\chi_1 g + \chi_2 \tilde{g}_1$. So, the metric is the product metric for $T-1 \leq  t_1 \leq T$ and our original metric for $T-3 \leq  t \leq T-2$. We extend $g_1$ to the rest of $M_1$ by letting $g_1=g$ for points in $M_1 \setminus \Sigma_1(T)$. Furthermore, notice that as $T\rightarrow \infty$, $g_1$ approaches $\tilde{g}_1$ uniformly in the region $T-2 \leq  t_1 \leq T-1$. We denote the cross section $\lbrace t_1=T \rbrace$ by $S_1$. 

We proceed inductively. Given $M_i$ as obtained above, we perform the same procedure on the $i+1$-th AC end. For each AC end we use the same parameter $T$ so we don't have to worry about keeping track of $n$ individual parameters representing how far down each cylindrical end we've performed the chopping. Thus, we obtain the manifold $M_n$ with a metric which we denote by $g_n$, and whose boundary consists of $\partial M \cup \left( \cup_{i=1}^n S_i \right)$. 

Now, we take two copies of $M_n$ which we denote by $M^+$ and $M^-$. We denote the surfaces $S_i$ in each of these sets by $S_i^+$ and $S_i^{-}$, respectively. We also denote the sets $\Sigma_i(T)$ in each $M^{\pm}$ by $\Sigma_i^{\pm}(T)$. We glue $M^+$ and $M^-$ by identifying the points of $S_i^+$ and $S_i^-$ via the identity map for $1\leq i \leq n$ to obtain a doubled manifold which we denote by $\hat{M}$.

	Consider the gluing map $G: M^+ \sqcup M^- \rightarrow \hat{M}$. For the $i$-th AC end we define the sets
	\begin{equation} \nonumber
	\Sigma_i(T)=\lbrace (t_i, \omega): T-3\leq t_i\leq T \rbrace, \quad \Gamma_i(\delta)=\lbrace (t_i, \omega): \delta \leq t_i\leq T \rbrace, \quad \Pi_i(\delta)=\lbrace (t_i, \omega): \delta < t_i\leq T \rbrace   
	\end{equation}
	Using natural identifications, we can consider these as subsets of $M^\pm$ and denote them by $\Sigma_i^\pm(T)$, $\Gamma_i^\pm(\delta), \Pi_i^\pm(\delta)$. We also define 
	\begin{gather} \nonumber
	\hat{\Sigma}_i^\pm(T)=G(\Sigma_i^\pm (T)), \quad \hat{\Gamma}_i^\pm(\delta)=G(\Gamma_i^\pm (\delta)), \quad \hat{\Pi}_i^\pm(\delta)=G(\Pi_i^\pm (\delta))  \\ \nonumber
	\hat{\Sigma}^\pm(T) = \cup_i \hat{\Sigma}_i^\pm(T), \quad \hat{\Gamma}^\pm(\delta) = \cup_i \hat{\Gamma}_i^\pm(\delta), \quad \hat{\Pi}^\pm(\delta) = \cup_i \hat{\Pi}_i^\pm(\delta) \\ \nonumber
	\hat{\Sigma}(T)= \hat{\Sigma}^+(T) \cup \hat{\Sigma}^-(T), \quad \hat{\Gamma}(\delta)= \hat{\Gamma}^+(\delta) \cup \hat{\Gamma}^-(\delta), \quad \hat{\Gamma}(\delta)= \hat{\Pi}^+(\delta) \cup \hat{\Pi}^-(\delta). 
	\end{gather}
	Furthermore, given some general set $P\subset M$ we similarly define corresponding sets $P^\pm \subset M^\pm$, $\hat{P}^\pm=G(P^\pm)$, $\hat{P}=\hat{P}^+ \cup \hat{P}^-$.

 Since the metrics $g^{\pm}$ agree on and are cylindrical near each $S_i^{\pm}$, $\hat{M}$ is a smooth Riemannian manifold with a metric which we denote by $\hat{g}$. It has two SAF ends. Since we can identify both $\hat{M}^\pm$ with the same subset of $M$, there is a natural correspondence between the points of $\hat{M}^+$ and $\hat{M}^-$. This allows us to define the following map. Given a point $x^+\in \hat{M}^+$ and the corresponding point $x^-\in \hat{M}^-$ we define the inversion map $I:\hat{M} \rightarrow \hat{M}$ by $I(x^{\pm})=x^{\mp}$. It is easy to see that this is an isometry of $(\hat{M}, \hat{g})$.

It is also easy to see that $\mathcal{F}=\cup_{i=1}^n \hat{S}^{\pm}_i$ is the fixed point set of this isometry. As a result, being totally geodesic, it is a minimal surface.  Hence, if we choose one of the SAF ends, which for definiteness we will say is the end contained in $\hat{M}^+$, there will be an outermost minimal surface with respect to this end, which we will denote by $\mathcal{S}$. We would now like to apply the KWY theorem to the region in $\hat{M}^+$ exterior to this minimal surface. However, currently our region does not necessarily satisfy all of the hypotheses of the theorem. We restore each of these hypotheses in turn.

\subsection{The Divergence Constraint} \label{DivergenceConstraint}

So far, we have not defined any electric and magnetic fields on $\hat{M}$. We can't simply push forward the $E^{\pm}$ and $B^{\pm}$ since they will not match across $\mathcal{F}$. The easiest way to sidestep this problem is to realize that we only need our vector fields to be defined and have the appropriate properties on $\hat{M}^+ \subset \hat{M}$. That is, we only need the vector fields to be defined, divergence free, and satisfy the charged dominant energy condition outside of the outermost minimal surface with respect to the $\hat{M}^+$ end in order to apply the KWY Theorem. Thus, we push forward the vector fields $E$ and $B$ using the inclusion map $\hat{M}^+ \xhookrightarrow[]{} \hat{M}$ to obtain vector fields $\hat{E}$ and $\hat{B}$ defined on $\hat{M}^+$ which, for the moment, are not divergence free with respect to the $\hat{g}$ metric. 

We will need elliptic estimates in oder to restore the divergence constraint and the CDEC by solving certain PDE. In order to do so, first we need to choose an open cover on $\hat{M}$ to define the weighted H\"{o}lder norms. We do this as follows. We have the SAF end $\Omega \subset M$ which we can identify with two subsets $\hat{\Omega}^+$ and $\hat{\Omega}^-$. Using the AC coordinates, each of the cylindrical necks $\hat{\Pi}_i (2)$ can be covered by sets $V_{i, 1}=(2, 2T-2)\times U_{i, 1}$ and $V_{i, 2}=2(2T-2)\times U_{i, 2}$ where $U_{i, 1}\cup U_{i, 2}=S^2$, each of which is diffeomorphic to $\tilde{V}_{i, k}=(2, 2T-2)\times D\subset \mathbb{R}^3$. The compact set $\hat{M}^+\setminus (\Omega^+\cup \hat{\Pi}(3))$ can be covered by finitely many pre-compact open sets diffeomorphic to standard balls and half balls (near $\partial \hat{M}$) in $\mathbb{R}^3$. We can also choose the coordinates in such a way that $\partial_{x^3}|_{x^3=0}=\partial_{\tau}$, the normal derivative with respect to the Riemannian metric induced on the half-ball from $(\hat{M}, \hat{g})$. We denote these sets by $\hat{W}^+_i$ with $1\leq i \leq m$ and their diffeomorphic images as $\tilde{W}^+_i \subset \mathbb{R}^3$. Then we can consider the sets $\hat{W}^-_i=I(\hat{W}^+_i)$ which then form a cover of the corresponding set in $\hat{M}^-$. 

All together the collection $\mathcal{C}=\lbrace \hat{\Omega}^+, \hat{\Omega}^-, V_{1, 1}, V_{1, 2}, \dots V_{n, 1}, V_{n, 2}, \hat{W}^+_1,  \hat{W}^-_1, \dots  \hat{W}^+_m, \hat{W}^-_m   \rbrace$ forms an open cover of $\hat{M}$ by sets diffeomorphic to subsets of $\mathbb{R}^3$. We then use these diffeomorphisms to define the weighted H\"{o}lder norms as in Section \ref{DefinitionHolderSpace}. 

We have the following proposition, which is a slight generalization of Proposition 1 in \cite{WY}, and follows the same proof. 

\begin{proposition} \label{EllipticEstimatesProposition}
Let $0<\beta<1, \nu > 2$, and let $h\in C^{0, \alpha}_{-\nu}$ satisfy $h\geq 0$. If $\phi \in \tilde{C}^{2, \alpha}_{-\beta}$ then there is a constant $C$ independent of $T$ such that for all sufficiently large $T$
\begin{equation} \label{EllipticEstimate}
\Vert \phi \Vert_{C^{2, \alpha}_{-\beta}} \leq C \left( \Vert\phi\Vert_{C^{0}_{-\beta}} + \Vert (\Delta_{\hat{g}}-h)\phi \Vert_{C^{0, \alpha}_{-\beta-2}}  \right). 
\end{equation}
\end{proposition}

\noindent \textbf{Proof:}
Consider the set $\Omega(1)$ (see above for the definition) and the corresponding sets $\hat{\Omega}(1)^\pm$ so that $\hat{\Omega}(1)^\pm\subset \hat{\Omega}^\pm$. In each of these ends we can use local estimates and the scaling of annuli as in Proposition 26 of \cite{SmithWeinstein} to obtain estimates
\begin{equation} \label{Prop1}
\Vert \phi \Vert_{C^{2, \alpha}_{-\beta}\left(\hat{\Omega}(1)^\pm\right)} \leq C \left( \Vert \phi \Vert_{C^{0}_{-\beta}\left(\hat{\Omega}^\pm\right)} + \Vert(\Delta_{\hat{g}}-h)\phi \Vert_{C^{0, \alpha}_{-\beta-2}\left(\hat{\Omega}^\pm\right)}  \right). 
\end{equation}

The set $K=\hat{M}\setminus\lbrace \hat{\Omega}(1)^+ \cup \hat{\Omega}(1)^-\cup \hat{\Pi}(4) \rbrace$ is compact. At the boundary we can take a cover by sets $U\subset U'$ such that in local coordinates these are diffeomorphic to Euclidean half balls of radius $1$ and $2$ (denoted $\bar{B}_1$ and $\bar{B}_2$)and such that $\partial_{\tau}=\partial_{x^3}|_{x^3=0}$ on the boundary portion. Using Lemma 6.29 and 6.35 in \cite{GT} (which is why we require the normal derivative at the boundary to vanish) we obtain estimates 
\begin{equation}
\Vert \phi \Vert_{C^{2, \alpha}(\bar{B}_1)} \leq C \left( \Vert \phi \Vert_{C^{0}(\bar{B}_{2})} + \Vert (\Delta_{\hat{g}}-h)\phi \Vert_{C^{0, \alpha}(\bar{B}_2)}   \right) 
\end{equation}     

Away from the boundary, we similarly can take a cover by open subsets $V\subset V'$ diffeomorphic to open Euclidean balls of radius $1$ and $2$ denoted by $B_1$ and $B_2$ to obtain estimates
\begin{equation}
\Vert \phi \Vert_{C^{2, \alpha}(B_1)} \leq C \left( \Vert \phi \Vert_{C^{0}(B_{2})} + \Vert (\Delta_{\hat{g}}-h)\phi \Vert_{C^{0, \alpha}(B_2)}   \right). 
\end{equation}      

We can choose the sets $U', V'$ in such a way that each one is a subset of at least one element of $\mathcal{C}$. Since each of the sets $\hat{W}^+_1,  \hat{W}^-_1, \dots  \hat{W}^+_m, \hat{W}^-_m  $ is precompact, we can take a finite open cover of $K$ using the sets of the type $U, V$ in such a way that it is also a cover of each of these sets. We use this same cover of $K$ for each $T$.    

The compact set $\hat{\Gamma}(3)$ can be covered by finitely many geodesic balls $B^{\hat{g}}_{q_i}(\rho)$ of radius $\rho > 0$ so that the elliptic constant of $\hat{g}$ when computed in geodesic normal coordinates on a geodesic ball $B^{\hat{g}}_{q_i}(2\rho)$ is uniformly bounded above and below. We can also choose these geodesic balls so that each one is a subset of at least one of the sets $V_{1, 1}, V_{1, 2}, \dots V_{n, 1}, V_{n, 2}$. The number of such balls increases as we increase $T$, but because the metric in each of the necks converges to the cylindrical metrics, $\rho$ and the bounds can be chosen independent of $T$. This yields local elliptic estimates
\begin{equation} \label{geodesicestimate}
||\phi||_{C^{2, \alpha}_{-\beta}(B^{\hat{g}}_{q_i}(\rho))} \leq C \left( ||\phi||_{C^{0}_{-\beta}(B^{\hat{g}}_{q_i}(2\rho))} + ||(\Delta_{\hat{g}}-h)\phi||_{C^{0, \alpha}_{-\beta-2}(B^{\hat{g}}_{q_i}(2\rho))}  \right). 
\end{equation}

Since the cover of $K$ is finite and the same for all $T$, and the elliptic constant in \eqref{geodesicestimate} can be chosen independent of $T$ we can put all of our estimates together to obtain 
\begin{equation} 
\Vert \phi \Vert_{C^{2, \alpha}_{-\beta}} \leq C \left( \Vert\phi\Vert_{C^{0}_{-\beta}} + \Vert (\Delta_{\hat{g}}-h)\phi \Vert_{C^{0, \alpha}_{-\beta-2}}  \right) 
\end{equation}
where the constant $C$ is independent of $T$. $\square$

\medskip

We also have a similar estimate for solutions of Poisson's equation on $\hat{M}^+$.

\begin{proposition} \label{PoissonEstimate}
	Let $0<\beta<1, \nu > 2$, and let $h\in C^{0, \alpha}_{-\nu}( \hat{M}^+)$ satisfy $h\geq 0$. If $\phi \in C^{2, \alpha}_{-\beta}( \hat{M}^+)$ and $\phi=0$ on $\partial \hat{M}^+$ then there is a constant $C$ independent of $T$ such that for all sufficiently large $T$
	\begin{equation} \label{PoissonEstimateEq}
	\Vert \phi \Vert_{C^{2, \alpha}_{-\beta}} \leq C \left( \Vert\phi\Vert_{C^{0}_{-\beta}} + \Vert (\Delta_{\hat{g}}-h)\phi \Vert_{C^{0, \alpha}_{-\beta-2}}  \right). 
	\end{equation}
\end{proposition}

\noindent \textbf{Proof:} The proof is the same as Proposition \ref{EllipticEstimatesProposition}, except that near the boundary of $\hat{M}^+$ we use Theorem 6.6 in \cite{GT}. $\square$

\medskip 

With this in hand, we can restore the divergence constraint.

\begin{proposition} \label{PoissonSolution}
	For each $T$ large enough, there exists a unique solution $\varphi\in C^{2, \alpha}_{-1}(\hat{M}^+)$ of the problem:
	\begin{equation}
	\Delta_{\hat{g}} \varphi =f, \quad \varphi|_{\partial \hat{M}^+}=0
	\end{equation}
	on $\hat{M}^+$, where $f=\div_{\hat{g}} \hat{E}$. Furthermore, 
	\begin{equation}
	||\varphi||_{C^{2, \alpha}_{-1}(\hat{\Omega}^+)} \leq \tilde{\varepsilon}(T)
	\end{equation}
	where $\lim_{T\rightarrow \infty} \tilde{\varepsilon}(T)=0$.
\end{proposition}

\begin{remark}
	This proposition corresponds to Proposition 2 in \cite{WY}. However, it contains a small gap which is easily remedied. The authors claim the existence of a $C^{2, \alpha}_{-1}$ solution by referring the reader to \cite{Choquet} and then use Proposition \ref{PoissonEstimate} to obtain the bound on the solution. However, both \cite{Choquet} and Proposition \ref{PoissonEstimate} apply in the case $0<\beta<1$. We will solve the problem with $\beta=2/3$ and then apply Lemma \ref{ADMestimate1} to obtain the $C^{2, \alpha}_{-1}$ bound.  
\end{remark}

\noindent \textbf{Proof:} The existence of a unique solution $\varphi \in C^{2, \alpha}_{-2/3}$ follows from \cite{Choquet}. The smallness of the solution in $C^{2, \alpha}_{-2/3}$ will follow from Proposition \ref{PoissonEstimate} once we obtain a $C^0_{-2/3}$ estimate. To do this, first we will establish an unweighted supremum bound by using the maximum principle. We will make use of the existence of a bounded subharmonic function $\varPsi$ on $\hat{M}^+$ which satisfies $\Delta_{\hat{g}} \varPsi > C > 0$ on $\hat{\Gamma}^+(T-2)$, and which is supported in $\hat{\Sigma}^+(T)$ with $C$ independent of $T$ for all $T$ large enough. The existence of this function is established in Proposition \ref{SubharmonicFunction}. 

Notice that $\Vert f \Vert_{C^{0, \alpha}_{-8/3}} = \varepsilon(T) \rightarrow 0$ as $T\rightarrow \infty$. Let $P=\sup_{\hat{M}^+} |\varPsi|$ (In fact $P=1$ as shown in Proposition \ref{SubharmonicFunction}). Then, by the above properties of $\varPsi$ the function $\varphi + \varepsilon(T) C^{-1} \varPsi$ satisfies $\Delta_{\hat{g}}(\varphi + \varepsilon(T) C^{-1} \varPsi)\geq 0$. Since $\varphi$ vanishes on $\partial \hat{M}^+$ and at $\infty$, by the maximum principle we have $\varphi+ \varepsilon(T) C^{-1} \varPsi \leq \varepsilon(T) C^{-1} P$, or $\varphi \leq 2\varepsilon(T) C^{-1} P$. Similarly, by considering the superharmonic function $\varphi - \varepsilon(T) C^{-1} \varPsi$ we obtain $\varphi \geq -2\varepsilon(T) C^{-1} P$, so that $\sup_{\hat{M}^+}|\varphi| \leq 2\varepsilon(T) C^{-1} P$.  

To obtain the weighted estimate, we take the SAF region $\hat{\Omega}^+$ and solve the problem
\begin{equation}
\Delta_{\hat{g}} v =0, \quad v=1 \; \text{on} \; \partial \hat{\Omega}^+, \quad v\rightarrow 0 \; \text{at} \; \infty.
\end{equation}
There is some constant $C_2$ such that $0<v<C_2 \sigma^{-1}$. Then the functions $\pm \varphi +2\varepsilon(T) C^{-1}Pv$ are harmonic in $\hat{\Omega}^+$, tend to $0$ at infinity, and are non-negative on $\partial \Omega$. By the maximum principle, $\pm \varphi +2\varepsilon(T) C^{-1}Pv\geq 0$ on $\Omega$, and so $|\varphi|\leq 2\varepsilon(T) C^{-1}Pv \leq 2\varepsilon(T) C^{-1}PC_2 \sigma^{-1}$ which yields $\sigma|\varphi|\leq C_3 \varepsilon(T)$. This implies $ \Vert \varphi \Vert_{C^0_{-2/3}}\leq C_3 \varepsilon(T)$. 

By Proposition \ref{PoissonEstimate} we have 
\begin{equation}
\Vert \varphi \Vert_{{C}^{2, \alpha}_{-2/3}} \leq C \left( C_3 \varepsilon(T) + \Vert f  \Vert_{C^{0, \alpha}_{-8/3}}  \right)\leq C_4 \varepsilon(T) 
\end{equation}
and by Lemma \ref{ADMestimate1}
\begin{equation} \label{chargeestimate}
\Vert \varphi \Vert_{{C}^{2, \alpha}_{-1}}(\hat{\Omega}^+) \leq CC_4\varepsilon(T)=\tilde{\varepsilon}(T).
\end{equation}
\noindent $\square$

\begin{remark}
	By elliptic regularity, the solution of Proposition \ref{PoissonSolution} is smooth on $\hat{M}^+$.
\end{remark}

We now establish the existence of the subharmonic function $\Psi$ used in the previous proposition. 

\begin{proposition} \label{SubharmonicFunction}
There exists a bounded subharmonic function $\varPsi$ on $\hat{M}^+$ such that $\Delta_{\hat{g}} \varPsi > C>0$ on $\hat{\Gamma}^+(T-2)$.
\end{proposition}

\noindent\textbf{Proof:}Pick one of the AC ends. Consider the region of $\hat{M}^+$ which corresponds to $T-4\leq t\leq T$. Define the function $s(t)$ (defined for $t\in \mathbb{R}$) by 
\begin{align*}
s(t)=      \begin{cases}
0, & t\leq 0 \\
e^{-1/t}, & t>0 
\end{cases}.	
\end{align*}  
Next, define the smooth function 
\begin{equation}
b(t)=\frac{s(t-(T-3))}{s(t-(T-3))+s((T-2)-t)}
\end{equation}   
which is $0$ for $T-4\leq t\leq T-3$, increasing on $T-3 < t < T-2$ and identically $1$ for $T-2\leq t \leq T$. Now, consider the function $k(t)=e^{\gamma (t-T)}$. Finally, consider the product function $b(t)k(t)$ for $T-4 \leq t \leq T$. This function is identically $0$ for $T-4\leq t \leq T-3$ and so we can extend it to be $0$ on the rest of $\hat{M}^+$. We claim that for sufficiently large $\gamma$ this is a subharmonic function. 

Recall that in local coordinates the Laplace-Beltrami operator takes the form 
\begin{equation}
\Delta_{\hat{g}}f = \frac{1}{\sqrt{|\hat{g}|}}\partial_i\left( \sqrt{|\hat{g}|} \hat{g}^{ij} \partial_j f \right)
\end{equation}
and if our function is only a function of $t$ then
\begin{equation}
\Delta_{\hat{g}}f = \frac{1}{\sqrt{|\hat{g}|}}\partial_i\left( \sqrt{|\hat{g}|} \hat{g}^{it} \partial_t f \right)=g^{tt}\partial_t^2f+\partial_i(g^{it})\partial_tf+ \frac{\partial_i (\sqrt{|\hat{g}|})}{\sqrt{|\hat{g}|}}g^{it}\partial_tf.
\end{equation}
Also, we have by the well known property of the Laplace-Beltrami operator: 
\begin{equation}
\Delta_{\hat{g}} (bk) = b\Delta_{\hat{g}}k + k\Delta_{\hat{g}} b +2\hat{g}(\nabla b, \nabla k).
\end{equation}
Since both $k$ and $b$ are nondecreasing functions of $t$, the term $\hat{g}(\nabla b, \nabla k)$ is nonnegative. Next, consider 
\begin{equation}
\Delta_{\hat{g}}k =\left(g^{tt}\gamma^2 +\partial_i(g^{it})\gamma + \frac{\partial_i (\sqrt{|\hat{g}|})}{\sqrt{|\hat{g}|}}g^{it}\gamma \right)e^{\gamma (t-T)}.
\end{equation}
Due to the cylindrical geometry, we have $g^{tt}\rightarrow 1$, $g^{it}\rightarrow 0$ for $i \neq t$, and $\partial_ig^{it}, \partial_t(\sqrt{|\hat{g}|})/\sqrt{|\hat{g}|} \rightarrow 0$ as $T \rightarrow \infty$. Hence, there will be some $T_{0}$ and $\gamma_0$ such that for all $T>T_{0}$ and all $\gamma>\gamma_0$, $\Delta_{\hat{g}} k> 0$ for $T-4\leq t \leq T$. Finally, let us look at the term 
\begin{equation}
\Delta_{\hat{g}}b =g^{tt}\partial_t^2b+\partial_i(g^{it})\partial_tb+ \frac{\partial_i (\sqrt{|\hat{g}|})}{\sqrt{|\hat{g}|}}g^{it}\partial_tb.
\end{equation}
Let us examine how this term behaves as we take $t\rightarrow T-3^+$. We see that for the leading terms we have 
\begin{equation}
\partial^2_tb \approx C_2 \frac{e^{-1/(t-T-3)}}{(t-(T-3))^4}
\end{equation}
and
\begin{align}
\partial_tb \approx C_3\frac{e^{-1/(t-(T-3))}}{(t-(T-3)^2}
\end{align}
where $C_2, C_3 >0$. And so, near $t=T-3$ we will have $\Delta_{\hat{g}} b \geq 0$. More specifically, there will be some $\epsilon$ (independent of $T$) such that $\Delta_{\hat{g}} b \geq 0$ on $[T-3, T-3+\epsilon]$. On $[T-3+\epsilon, T]$ we can write

\begin{equation}
b\Delta_{\hat{g}}k + k\Delta_{\hat{g}} b=\left(bg^{tt}\gamma^2 +\partial_i(g^{it})(b\gamma +\partial_tb )+ \frac{\partial_i (\sqrt{|\hat{g}|})}{\sqrt{|\hat{g}|}}g^{it}  (b\gamma+\partial_tb) +g^{tt}\partial_t^2b \right)e^{\gamma (t-T)}
\end{equation}
Thus, since $b(t)\geq b(T-3+\epsilon)>0   $ on $[T-3+\epsilon, T]$, we see that for sufficiently large $\gamma$, $b\Delta_{\hat{g}}k + k\Delta_{\hat{g}} b >0$, and so there will be some constant $C$ such that $b\Delta_{\hat{g}}k + k\Delta_{\hat{g}} b >C$ on $\hat{\Gamma}^+(T-2)$. We construct such functions in each of the cylindrical ends, and then take their sum, defining it to be $\varPsi$, which is thus the required bounded subharmonic function.$\qed$

\medskip

We define $E'=E-\nabla_{\hat{g}}\varphi$, which is now a smooth, divergence free vector field on $\hat{M}^+$. Using \eqref{charges} we compute the accompanying charge
\begin{equation}
|q_{e, T}-q_e|= \left|\lim_{r\rightarrow \infty} \frac{1}{4\pi} \int_{S_r} (\nabla_{\hat{g}} \varphi)_j \nu^j dS \right| \leq C\tilde\varepsilon(T)
\end{equation}
for some constant $C$ independent of $T$. We similarly obtain the vector field $B'$ and conclude that $|q_{b, T}-q_b|\leq C \tilde{\varepsilon}(T)$. Therefore, if we denote the total squared charge by $q_T^2$, we see that $\lim_{T\rightarrow \infty} q^2_T = q^2$. 

\subsection{The CDEC Constraint} \label{CDECConstraint}

We are looking for a conformal deformation $\bar{g}=u^4 \hat{g}$ on $\hat{M}$ such that the scalar curvature $\bar{R}$ of this metric satisfies the charged dominant energy condition. Recall that by Proposition \ref{ConformalDeformation1}, we can assume that our original data set satisfies the strict charged dominant energy condition. Given the vector fields $E', B'$ we define $\bar{E}=u^{-6}E', \bar{B}=u^{-6}B'$ which are defined and divergence free on $\hat{M}^+$. In order to restore the charged dominant energy condition on $\hat{M}^+$ we want 
\begin{equation}
\bar{R} \geq 2(|\bar{E}|^2_{\bar{g}}+|\bar{B}|^2_{\bar{g}})=2u^{-8}\left(|E'|^2_{\hat{g}}+|B'|^2_{\hat{g}}\right).
\end{equation}  
We begin with the following Lemma, whose importance will soon become apparent. Recall the inversion map defined previously by $I(x^{\pm})=x^{\mp}$.

\begin{lemma} \label{zetaLemma}
	For each $T$ there exists a function $\zeta\geq 0$ on $\hat{M}$ which satisfies 
	
	\begin{itemize} 
		\item[(i)] $\hat{R}+\zeta\geq 2\left( |E'|^2_{\hat{g}} + |B'|^2_{\hat{g}}    \right) \; \text{on} \; \hat{M}^+$
		
		\item[(ii)]$\zeta \; \text{is smooth and }\zeta(x)=\zeta(I(x))$ 
		
		\item[(iii)] $4\hat{R}+3\zeta > 0 $
		\item[(iv)] $\Vert \zeta \Vert_{C^{0, \alpha}_{-8/3}} \rightarrow 0 \quad \text{as} \quad T\rightarrow \infty$. 
	\end{itemize}
	
\end{lemma}

\noindent \textbf{Proof:} We can write (i) as
\begin{align} \nonumber
\hat{R}+\zeta
\geq 2\left(|E'|^2_{\hat{g}} +  |B'|^2_{\hat{g}}\right)= 2\left(|\hat{E}|^2_{\hat{g}} +  |\hat{B}|^2_{\hat{g}}\right) + 4\langle \hat{E}, \nabla \varphi_E \rangle + 4\langle \hat{B}, \nabla \varphi_B \rangle + |\nabla \varphi_E|^2 + |\nabla \varphi_B|^2
\end{align}
or
\begin{equation} \nonumber
\left[ \hat{R} - 2\left(|\hat{E}|^2_{\hat{g}} +  |\hat{B}|^2_{\hat{g}}\right)  \right] + \zeta - \left( 4\langle \hat{E}, \nabla \varphi_E \rangle + 4\langle \hat{B}, \nabla \varphi_B \rangle + |\nabla \varphi_E|^2 + |\nabla \varphi_B|^2  \right) \geq 0
\end{equation}.

On $\hat{M}^+$ outside of $\hat{\Gamma}^+(T-2)$ the first term is positive by the strict CDEC. In $\hat{\Gamma}^+(T-2)$  it satisfies $\hat{R} - 2\left(|\hat{E}|^2_{\hat{g}} +  |\hat{B}|^2_{\hat{g}}\right) \geq -\varepsilon(T)$ for some $\varepsilon(T)>0$ such that $\varepsilon(T)\rightarrow 0$ as $T\rightarrow \infty$. Therefore, we look for a smooth function such that  
\begin{equation} \label{zeta}
\zeta\geq \vert 4\langle \hat{E}, \nabla \varphi_E \rangle + 4\langle \hat{B}, \varphi_B \rangle + |\nabla \varphi_E|^2 + |\nabla \varphi_B|^2 \vert
\end{equation}
everywhere on $\hat{M}^+$ and 
\begin{equation}
\zeta \geq 2\varepsilon(T) + \vert 4\langle \hat{E}, \nabla \varphi_E \rangle + 4\langle \hat{B}, \varphi_B \rangle + |\nabla \varphi_E|^2 + |\nabla \varphi_B|^2 \vert
\end{equation}
on $\hat{\Gamma}^+(T-2)$. We require $2\varepsilon(T)$ in order for (iii) to hold. Obviously we can find such a function satisfying (ii), and (iv) follows since $\Vert \varphi_E \Vert_{{C}^{2, \alpha}_{-1}}, \Vert \varphi_B \Vert_{{C}^{2, \alpha}_{-1}} \rightarrow 0$ as $T\rightarrow \infty$ so we can choose $\zeta$ to fall-off rapidly at infinity. In fact, since $E, B$ are $O(1/|x|^2)$ we can obtain $\Vert \zeta \Vert_{C^{0, \alpha}_{-4}}\rightarrow 0$ as $T\rightarrow \infty$. $\square$

With $\zeta$ in hand we seek to solve the problem $\bar{R}=(\hat{R}+\zeta)u^{-8}$, $\partial_{\tau} u =0$ which by using \ref{ScalarUnderConformalChange} can be written as 
\begin{align} \begin{split} \label{equation}
\Delta_{\hat{g}}u-\frac{1}{8}\left( \hat{R}u-\frac{\hat{R}+\zeta}{u^3}   \right)=0\\
\partial_{\tau} u= 0 \quad \text{on} \quad \partial \hat{M}.
\end{split}
\end{align}

\begin{proposition} \label{UniformBound}
	Let 
	\begin{equation}
	h=\frac{1}{2}\hat{R}+\frac{3}{8}\zeta.
	\end{equation}
	Then there is a constant $C$ independent of $T$, for $T$ large enough, such that if $\phi \in \tilde{C}^{2, \alpha}_{-2/3}$ then 
	\begin{equation}
	||\phi||_{C^{2, \alpha}_{-2/3}} \leq C|| (\Delta_{\hat{g}}-h) \phi ||_{C^{0, \alpha}_{-8/3}}.
	\end{equation}
\end{proposition}

\noindent \textbf{Proof:} The proof is nearly identical to that of the corresponding Proposition 3 in \cite{WY}, and is hence omitted. We only remark that the proof relied on the fact that the scalar curvature of Majumdar-Papapetrou data was non-negative. Thus we needed the results of Section \ref{InitialConformalDeformation} for the proof to carry over. $\square$

\medskip

\begin{proposition}
	There exists a unique solution $u=1+\psi$ to \eqref{equation} such that $\Vert \psi  \Vert_{\tilde{C}^{2, \alpha}_{-2/3}} \rightarrow 0$ as $T\rightarrow \infty$. 
\end{proposition}

\textbf{Proof:} We follow the procedure and notation outlined in \cite{WY}. We look for a conformal perturbation $u=1+\psi$, in which case \eqref{equation} takes the form
\begin{equation} \label{equation2}
\Delta_{\hat{g}}\psi-\frac{1}{8}\left( \hat{R}\psi-\frac{\hat{R}+\zeta}{(1+\psi)^3}   \right)-\frac{1}{8}\hat{R}=0.
\end{equation}

Therefore we define an operator $\mathcal{N}: \tilde{C}^{2, \alpha}_{-2/3}\rightarrow C^{0, \alpha}_{-8/3}$ by
\begin{equation}
\mathcal{N}(\psi)=\Delta_{\hat{g}}\psi-\frac{1}{8}\left( \hat{R}\psi-\frac{\hat{R}+\zeta}{(1+\psi)^3}   \right)-\frac{1}{8}\hat{R}
\end{equation}
Computing the Frechet derivative at $\psi=0$ we obtain
\begin{equation}
d\mathcal{N}=\Delta_{\hat{g}}-\frac{1}{8}\left(4\hat{R}+3\zeta \right)=\Delta_{\hat{g}}-h: \; \tilde{C}^{2, \alpha}_{-2/3}\rightarrow C^{0, \alpha}_{-8/3}.
\end{equation}
By our choice of $\zeta$, $4\hat{R}+3\zeta > 0$, and so by the Fredholm alternative, this operator is invertible, that is there exists $d\mathcal{N}^{-1}: C^{0, \alpha}_{-8/3} \rightarrow \tilde{C}^{2, \alpha}_{-2/3}$. By Proposition \ref{UniformBound} this operator is uniformly bounded, that is, there exists a constant $C$ independent of $T$ such that 
\begin{equation}
|| d\mathcal{N}^{-1} x ||_{\tilde{C}^{2, \alpha}_{-2/3}} \leq C || x ||_{C^{0, \alpha}_{-8/3}}.
\end{equation}  

We define the quadratic part of $\mathcal{N}$ by
\begin{equation}
\mathcal{Q}(\psi) \equiv \mathcal{N}(\psi)-\mathcal{N}(0)-d\mathcal{N}(\psi)
\end{equation}
and a subsequent calculation shows
\begin{equation}
\mathcal{Q}(\psi)=\frac{\hat{R}+\zeta}{8}\left( \frac{6+8\psi+3\psi^2}{(1+\psi)^3}     \right)\psi^2.
\end{equation}
Hence, there exists some $\eta_0$ such that for $||\psi||_{\tilde{C}^{2, \alpha}_{-2/3}} < \eta < \eta_0$
\begin{align}
&||\mathcal{Q}(\psi)||_{C^{0, \alpha}_{-8/3}}\leq C\eta^2 \\
&||\mathcal{Q}(\psi_1)-\mathcal{Q}(\psi_2)||_{C^{0, \alpha}_{-8/3}}\leq 2C\eta ||\psi_1-\psi_2||_{\tilde{C}^{2, \alpha}_{-2/3}}
\end{align}
where $C>0$ is some constant. The key to notice is that for all $\zeta$ with $\Vert \zeta \Vert_{C^{0, \alpha}_{-8/3}} <C_1$ we can choose $C$ to be independent of $\zeta$.   

Next, notice 
\begin{equation}
||\mathcal{N}(0)||_{C^{0, \alpha}_{-8/3}}=\bigg |\bigg |\frac{\zeta}{8}\bigg | \bigg |_{C^{0, \alpha}_{-8/3}} \leq \varepsilon(T)
\end{equation}
where $\varepsilon(T)\rightarrow 0$ as $T\rightarrow \infty$. 

Now choose $0<\lambda <1$ and $\eta >0$ such that 
\begin{align*}
\eta< \frac{\lambda}{2C^2}
\end{align*}
and $T$ large enough so that 
\begin{align*}
\varepsilon(T) < C\eta^2.
\end{align*}
Notice that the larger we take $T$, the smaller $\varepsilon(T)$ becomes, and hence the smaller we can choose $\eta$.

Now, consider the operator 
\begin{equation}
F(\psi)\equiv -d\mathcal{N}^{-1}\left( \mathcal{N}(0)+\mathcal{Q}(\psi)  \right)
\end{equation}
and notice 
\begin{align*}
||{F}(\psi)||_{\tilde{C}^{2, \alpha}_{-2/3}} & \leq C\left( ||\mathcal{N}(0)||_{C^{0, \alpha}_{-8/3}} + ||\mathcal{Q}(\psi)||_{C^{0, \alpha}_{-8/3}}   \right) \nonumber \\ & \leq C(\varepsilon(T)  + C\eta^2) \leq 2C^2\eta^2 < \frac{2C^2\lambda}{2C^2}\eta < \eta
\end{align*}
while
\begin{align*}
||{F}(\psi_1)-{F}(\psi_2)||_{\bar{C}^{2, \alpha}_{-2/3}} & \leq C||\mathcal{Q}(\psi_1)-\mathcal{Q}(\psi_2)||_{C^{0, \alpha}_{-8/3}} \\ & \leq 2C^2\eta ||\psi_1 - \psi_2||_{\tilde{C}^{2, \alpha}_{-2/3}} \\ & < \lambda ||\psi_1 - \psi_2||_{\tilde{C}^{2, \alpha}_{-2/3}}.
\end{align*}

Therefore, ${F}$ is a contraction mapping of the ball of radius $\eta$ to itself, and hence possesses a unique fixed point, which we continue to denote by $\psi$. Since
\begin{equation}
{F}(\psi)=\psi=-d\mathcal{N}^{-1}(\mathcal{N}(0)+\mathcal{Q}(\psi))
\end{equation}  
then operating by $d\mathcal{N}$ and rearranging 
\begin{equation}
d\mathcal{N}(\psi) +\mathcal{N}(0)+\mathcal{Q}(\psi)=\mathcal{N}(\psi)=0
\end{equation}
and so $u=1+\psi$ is precisely the conformal factor we were looking for. Furthermore $||\psi||_{\tilde{C}^{2, \alpha}_{-2/3}} < \eta \rightarrow 0$ as $T\rightarrow \infty$. $\square$

\begin{cor}
	The conformal factor $u$ satisfies $u(x)=u(I(x))$. 
\end{cor}   

\noindent \textbf{Proof:} Define $\tilde{u}(x)=u(I(x))$. Since $\hat{R}(x)=\hat{R}(I(x))$ and $\zeta(x)=\zeta(I(x))$, $\tilde{u}$ satisfies \eqref{equation}. Since the solution $u$ was obtained using the contraction mapping principle, it is unique, and so $\tilde{u}=u$. $\square$

\begin{remark}
By elliptic regularity, the above solution is smooth.
\end{remark}

\begin{cor}
The surface $\mathcal{F} \subset \hat{M}$ is a minimal surface of $(\hat{M}, \bar{g})$ where $\bar{g}=u^4\hat{g}$.
\end{cor}

\noindent \textbf{Proof:} The map $I$ remains an isometry of with respect to the new metric and $\mathcal{F}$ is its fixed point set. $\qed$

\medskip

To finish our proof, we need to obtain a bound on $\Vert \phi \Vert_{C^{2, \alpha}_{-1}}$ in order to have $\bar{m}$ close to $m$. As remarked earlier, we only need this in the SAF region $\hat{\Omega}^+$. Notice
\begin{equation}
\Delta_{\hat{g}}\psi=\frac{1}{8} \hat{R}\psi-\frac{\hat{R}+\zeta}{8(1+\psi)^3}   +\frac{1}{8}\hat{R}.
\end{equation}
Consider the right hand side as a function on $\hat{\Omega}^+$. We first show that the right hand side tends to $0$ in $C^{0, \alpha}_{-3}(\hat{\Omega}^+)\cap L^1(\hat{\Omega}^+)$ as $T\rightarrow \infty$. We rewrite the right hand side as 
\begin{equation}
\frac{1}{8} \hat{R}\psi-\frac{\zeta}{8(1+\psi)^3}   +\frac{\hat{R}}{8}\frac{(1+\psi)^3-1}{(1+\psi)^3}=\frac{1}{8} \hat{R}\psi-\frac{\zeta}{8(1+\psi)^3}   + \frac{\hat{R}}{8} \frac{3\psi+3\psi^2+\psi^3}{(1+\psi)^3}.
\end{equation}
We use the fact that if $f_i \in C^{0, \alpha}_{-\beta_i}, i=1,2$ and $\beta_1+\beta_2>3$, then
\begin{equation}
\Vert f_1f_2 \Vert_{C^{0, \alpha}_{-3}\cap L^1}=\Vert f_1f_2\Vert_{C^{0, \alpha}_{-3}} + \Vert f_1f_2 \Vert_{L^1}\leq C \Vert f_1 \Vert_{C^{0, \alpha}_{-\beta_1}   } \Vert f_2 \Vert_{C^{0, \alpha}_{-\beta_2}   }
\end{equation}
and so if one of the terms on the right-hand side is bounded and the other tends to $0$, the left-hand side tends to $0$ as well. By our previous remarks, we have that $\zeta \in C^{0, \alpha}_{-4}$ with $\Vert \zeta \Vert_{C^{0, \alpha}_{-4}}\rightarrow 0$ as $T\rightarrow \infty$, while the other two terms are of the form $f\psi$ with $f\in C^{0, \alpha}_{-3}$ bounded, and $\Vert \psi \Vert_{C^{0, \alpha}_{-2/3}} \rightarrow 0.$ We conclude that 
\begin{equation}
\Vert \Delta_{\hat{g}} \psi \Vert_{C^{0, \alpha}_{-3}\cap L^1} \rightarrow 0.
\end{equation}
Therefore, we apply Lemma \ref{ADMestimate1} to conclude 
\begin{equation} \label{MassBound}
\Vert \psi \Vert_{C^{2, \alpha}_{-1}} \rightarrow 0 \quad  \text{as} \quad T\rightarrow \infty.
\end{equation}

\subsection{Proof of Theorem 1: Part 1} \label{Proofof1Part1}

Consider the manifold $(\hat{M}, \bar{g})$ and notice that $\mathcal{F} \cup \partial \hat{M}^+$ is a minimal surface. Therefore, with respect to our chosen asymptotically flat end (that is, the end on which $\bar{E}$ and $\bar{B}$ are defined) there is some outermost minimal surface which we denote by $\mathcal{S}_{T}$, to show the dependence on $T$. Now, we consider the areas of this surface measured in the $\bar{g}$, $\hat{g}$ and $g$ metrics. We will also write $m_T$ and $|q_T|$ to show the dependence of the mass and charge on $T$.  

Of course, as we increase $T$ and move down the cylindrical ends, the surface $\mathcal{S}_{T}$ changes. However, as we do so, $\hat{g}\rightarrow g$, and $\bar{g}\rightarrow \hat{g}$ since $\psi \rightarrow 0$. Hence, we obtain an inequality of the form 
\begin{equation}
(1+5\varepsilon(T))|\mathcal{S}_{T}|_g\geq |\mathcal{S}_{T}|_{\bar{g}}=4\pi \rho_{T}^2\geq(1-5\varepsilon(T))|\mathcal{S}_{T}|_g>(1-5\varepsilon(T))A=4\pi(1-5\varepsilon(T))\rho^2.
\end{equation}
where $\varepsilon(T)\rightarrow \infty$ as $T\rightarrow \infty$.

On the other hand, the mass and charge of our chosen end satisfy
\begin{equation}
|m_T-m| < C\varepsilon(T)
\end{equation}
and 
\begin{equation}
\vert |q_T|-|q| \vert < C\varepsilon(T)
\end{equation}
for some $C$ independent of $T$ by \eqref{ADM} and \eqref{MassBound}.

If we denote $\lim_{T \rightarrow \infty} \rho_{T} = \bar{\rho}$ (which exists, after potentially passing to a subsequence)  we have by Corollary 1.2 of \cite{KhuriWeinsteinYamada2} that 

\begin{equation}
\rho_{T}\leq m_{T} + \sqrt{m_{T}^2-q_{T}^2}
\end{equation} 
and now, taking the limit and noting $\rho \leq \bar{\rho}$ we obtain 
\begin{equation}
\rho \leq \bar{\rho} \leq m+\sqrt{m^2-q^2}
\end{equation}
proving the upper bound in \eqref{penrose inequalities}.

As for \eqref{charged penrose}, suppose $\rho \geq |q|$. Take any $0<\lambda<1$ and instead of taking our initial data set $(M, g, E, B)$ consider $(M, g, \lambda E, \lambda B)$ which has charge $\lambda |q|$. Therefore $\rho > \lambda |q|$ (if $|q|=0$ this is trivially tue). In that case, eventually the $\rho_{T}$ satisfy $\rho_{T}>\lambda|q_{T}|$ and thus by Theorem 1.1 of \cite{KhuriWeinsteinYamada2} the solutions satisfy
\begin{equation} \label{cpiphi}
m_{T} \geq \frac{1}{2}\left( \rho_{T} + \frac{\lambda^2 q_{T}^2}{\rho_{T}}   \right).
\end{equation}
Taking the limit, we obtain
\begin{equation}
m \geq \frac{1}{2}\left( \bar{\rho} + \frac{\lambda^2 q^2}{\bar{\rho}}   \right) \geq  \frac{1}{2}\left( \rho + \frac{\lambda^2 q^2}{\rho}   \right)
\end{equation}
where the last inequality follows from $\bar{\rho}\geq \rho >\lambda|q|$. Since this holds for arbitrary $\lambda$, taking the limit $\lambda \rightarrow 1$ we are done.

Therefore we have proven our theorem in the case that our initial data satisfies the strict charged dominant energy condition and the outer-minimizing generalized boundary condition. For the case where the data merely satisfies the charged dominant energy condition, we apply Proposition \ref{ConformalDeformation1} to obtain a data set satisfying the strict charged dominant energy condition with mass $m_{\varepsilon}$ and the same charge $|q|$, with $m_\varepsilon \rightarrow m$ as $\varepsilon \rightarrow 0$. By \eqref{40} we have that any surface $S$ enclosing the generalized boundary satisfies $|S|>4\pi \rho_{\varepsilon}$ where $\rho_\varepsilon \rightarrow \rho$ as $\varepsilon \rightarrow 0$. Therefore the data satisfies

\begin{equation}
\rho_{\varepsilon} \leq m_{\varepsilon}+\sqrt{m_{\varepsilon}^2 - q^2}
\end{equation}
for all $\varepsilon$ and taking the limit $\varepsilon \rightarrow 0$ we have obtained 
\begin{equation}
\rho \leq m + \sqrt{m^2-q^2}.
\end{equation}

For the full Penrose inequality assume $\rho \geq |q|$ and choose any $\lambda$ with $0<\lambda<1$. Then for all sufficiently small $\varepsilon$ we have $\rho_{\varepsilon} \geq \lambda |q|$ and so
\begin{equation}
m_\varepsilon \geq  \frac{1}{2}\left( \rho_\varepsilon + \frac{\lambda^2 q^2}{\rho_\varepsilon}   \right)
\end{equation}
and taking the limit $\varepsilon \rightarrow 0$ we obtain
\begin{equation}
m \geq  \frac{1}{2}\left( \rho + \frac{\lambda^2 q^2}{\rho}   \right)
\end{equation}
and finally letting $\lambda \rightarrow 1$ we obtain the full charged Penrose inequality. $\square$

\subsection{The Case of Equality}

\label{Case of Equality}

First we remark that equality in the upper bound in \eqref{penrose inequalities} is equivalent to equality in \eqref{charged penrose}. Next we apply the argument of Section 7 of \cite{KhuriWeinsteinYamada2} to conclude that if equality occurs, our initial data must have either a single boundary component or a single asymptotically cylindrical end. The case of a single boundary component is treated in \cite{DisconziKhuri}. Thus, we are left with the case of a single asymptotically cylindrical end. We will use the same strategy as in \cite{DisconziKhuri}, which depends on the existence of a solution to the IMCF. 
\begin{proposition} \label{ExistenceIMCF}
	Suppose our initial data set has a single AC end. Then there exists a solution of the IMCF on all of $M$.  
\end{proposition}

\noindent \textbf{Proof:} Let $(t, \omega)\in [0, \infty) \times S^2$ be coordinates on the AC end (recall that by our convention $t\rightarrow \infty$ corresponds to moving down the AC end). Take any cross section $S_n$ with $n \in \mathbb{N}$. By Theorem 3.1 in \cite{HuiskenIlmanen}, there exists a solution $u_n$ on $\Omega_n=M\setminus E_n$ which satisfies the gradient estimate
\begin{equation} \label{gradient estimate}
|\nabla u_n (x)| \leq \sup_{\partial E_n \cap B_r(x)} H_+ + \frac{C}{r} \quad \text{a.e.}\quad x\in \Omega_n
\end{equation} 
for each $0<r\leq \sigma (x)$. Here, $H_+=\max (0, H)$, $E_n=\lbrace (t, \omega): t>n\rbrace$, $\sigma(x)$ is a positive, continuous function of $x$, and $C$ is a constant depending only on the dimension of $M$. 

Consider this sequence of solutions. We have $u_n=0$ for $x\in E_n=\lbrace (t, \omega): t>n\rbrace$ and so in order to obtain convergence, we need to normalize the functions. This is easy to do. Pick any point $x_0\in M$. Now, define new functions $\tilde{u}_n(x)=u_n(x)-u_n(x_0)$ so that all of these functions agree at the point $x_0$ and $\nabla u_n = \nabla \tilde{u}_n$. For simplicity we will subsequently drop the tilde and refer to these normalized functions as $u_n$.

As mentioned before, $\sigma(x)$ is a continuous function (for its full definition we refer the reader to \cite{HuiskenIlmanen}), so around every $x$ we can find an open neighborhood $B_x$ diffeomorphic to a ball such that for all $y\in B_x$ we have $\sigma(y)\geq \frac{1}{2}\sigma(x)$. Notice that as we take cross sections further down the cylindrical end, $H_{+}$ tends to $0$. Thus, for every $B_x$ we can find a constant $K_x$ such that $|\nabla u_n(y)|\leq K_x$ for a.e $y\in B_x$, so the family $\lbrace u_n \rbrace$ is equicontinuous in $B_x$. 

Now, take any point $x$ and take a finite length smooth curve with length $L$ joining it to $x_0$. By compactness, we can cover the curve with finitely many open balls, and in each ball we have a constant for the gradient as above. This shows that the family of functions is pointwise bounded.   

By the Arzela-Ascoli theorem, the sequence $\lbrace u_n \rbrace$ has a subsequence which converges uniformly on compact subsets to a function $u$. By Theorem 2.1 in \cite{HuiskenIlmanen}, $u$ is a solution of the weak IMCF. $\square$

\medskip

Denote $S_{i, \tau}=\lbrace x: u_i(x) = \tau \rbrace$ and $ S_\tau=\lbrace x: u(x) = \tau \rbrace$. By the remarks following Theorem 2.1 in \cite{HuiskenIlmanen} (which themselves depend on the Regularity Theorem 1.3.ii in this same paper) we have that 
\begin{equation} \label{IMCFconvergence}
S_{i, \tau} \rightarrow S_{\tau} \quad \text{locally in } C^{1, \alpha}.  
\end{equation}  
Furthermore, the flow has no jump discontinuities (as that would indicate the presence of another minimal surface, contradicting equality).

The key property of weak IMCF is the monotonicity of the charged Hawking mass. The charged Hawking mass is defined by
\begin{equation} \label{CHM2}
M_{CH}(S)=\sqrt{\frac{|S|}{16\pi}}\left( 1 + \frac{4\pi q^2}{|S|} - \frac{1}{16\pi} \int_{S} H^2  \right).
\end{equation}  
If we denote the surfaces of the IMCF by $S_{\tau}$ then the charged Hawking mass satisfies the well known property
\begin{align} \begin{split} \label{DerivativeCHM}
\frac{d}{d\tau} M_{CH}(S_{\tau})=-\frac{1}{2}\sqrt{\frac{\pi}{|S_{\tau}|}}q^2 + \sqrt{\frac{|S_{\tau}|}{16\pi}} \left( \frac{1}{2} - \frac{1}{4} \chi(S_\tau)  \right) \\
+ \frac{1}{16\pi} \sqrt{\frac{|S_{\tau}|}{16\pi}} \int_{S_{\tau}} \left(  2 \frac{|\nabla_{S_{\tau}} H |^2}{H^2} + |\Pi|^2 -\frac{1}{2}H^2 + R    \right)
\end{split}
\end{align}
which holds in the appropriate weak sense. Here $\Pi$ denotes the (weak) second fundamental form \cite{HuiskenIlmanen}, \cite{DisconziKhuri}.

\medskip

\noindent \textbf{Proof of Theorem 1: The Case of Equality} Assume equality holds in \ref{charged penrose}. Take the $u$ solving the weak IMCF which we constructed earlier. We point out a few facts about the resulting flow.

Our solution $u$ only takes values $\tau \in (c, \infty)$ where $c$ is some finite real number. This can be seen by combining the well known exponential growth condition for the areas of the flow \cite{HuiskenIlmanen} 
\begin{equation}
|S_{\tau}|=|S_{\tau_0}|e^{\tau-\tau_0}
\end{equation}
and the fact that by assumption any surfaces enclosing the cylindrical end have area greater than $A$. Therefore by normalizing we can assume $\tau \in (0, \infty)$. 

It is then easy to see that
\begin{equation*}
	\lim_{\tau \rightarrow 0^+} |S_\tau|=A
\end{equation*}  
and
\begin{equation*}
\lim_{\tau \rightarrow 0^+} \int_{S_{\tau}} H^2 =0.
\end{equation*}  
Next we argue that the flow is smooth for all $\tau$. The argument is a verbatim repetition of the argument in Section 8 of \cite{HuiskenIlmanen}. By the CDEC \eqref{DerivativeCHM} is greater than or equal to $0$. To see this, notice that by H\"older's inequality and the charged dominant energy condition
\begin{align} \label{Holder}
\begin{split}16\pi^2 q^2 = \left( \int_{S_{\tau}} E_j \nu^j  \right)^2 + \left( \int_{S_{\tau}} B_j \nu^j  \right)^2  \leq \left( \int_{S_{\tau}} |E_j \nu^j|  \right)^2 + \left( \int_{S_{\tau}} |B_j \nu^j|  \right)^2 \\
\leq \left( \int_{S_{\tau}} |E|  \right)^2 + \left( \int_{S_{\tau}} |B|  \right)^2 \leq |S_{\tau}| \int_{S_{\tau}} (|E|^2+|B|^2) \leq \frac{|S_{\tau}|}{2} \int_{S_{\tau}} R 
\end{split}
\end{align} 
which implies that 
\begin{equation}
-\frac{1}{2}\sqrt{\frac{\pi}{|S_\tau}|}q^2 + \frac{1}{16\pi}\sqrt{ \frac{|S_\tau|}{16\pi}  } \int_{S_\tau} R \geq 0. 
\end{equation}
On the other hand, $|\Pi|^2-(1/2)H^2$ is nonnegative (in the weak sense) \cite{HuiskenIlmanen}, while the remaining terms are nonnegative. 

Hence in order for equality to occur, \eqref{DerivativeCHM} must equal $0$ for almost every $\tau$. By Lemma 5.1 in \cite{HuiskenIlmanen} we have $H>0$ a.e. on on $S_\tau$ for a.e. $\tau$, and so $\int_{S_{\tau}} |\nabla_{S_{\tau}} H|^2 =0$ for a.e. $\tau$ and therefore by (1.10) in \cite{HuiskenIlmanen} and lower semicontinuity
\begin{equation}
\int_{S_{\tau}} |\nabla_{S_{\tau}} H|^2 =0 \quad \text{for all} \quad \tau
\end{equation}
Therefore
\begin{equation}
H_{S_\tau}(x)=H({\tau}) \quad \text{for a.e.} \; x\in S_{\tau}, \quad \text{for all} \; \tau \geq 0
\end{equation}
so that each $S_{\tau}$ has constant mean curvature. Since $H$ is locally bounded and $S_{\tau}$ has locally uniform $C^1$ estimates, it follows by elliptic theory that each $S_{\tau}$ is smooth. Furthermore, the flow does not jump at any $\tau$ since that would contradict the assumption that our initial data set does not contain any compact minimal surfaces. Hence, $H>0$ for $\tau>0$ which implies $H(\tau)$ is locally uniformly positive for $\tau>0$.

By Lemma 2.4 of \cite{HuiskenIlmanen}, for each $s>0$ there is some maximal $T$ such that the flow $(S_{\tau})_{s\leq \tau \leq T}$ is smooth. By the above regularity, $S_\tau$ has uniform space and time derivatives as $\tau \nearrow T$ and so the evolution can be smoothly continued past $\tau=T$. Hence $T=\infty$ and the flow is smooth everywhere. 

The next part follows from the arguments in \cite{DisconziKhuri} (see also \cite{Abramovic, KhuriWeinsteinYamada2}). Since the flow is smooth and \eqref{DerivativeCHM} vanishes we must have that equality holds everywhere in \eqref{Holder}. In particular
\begin{equation}
R=2(|E|^2+|B|^2), \quad E_j\nu^j=constant, \quad B_j\nu^j=constant, \quad E=e(\tau) \nu, \quad B=b(\tau) \nu 
\end{equation}
on each $S_{\tau}$. Furthermore, since $|\Pi|^2-\frac{1}{2}H^2=\left(  \lambda_1 - \lambda_2 \right)^2$, we must have that $\lambda_1 = \lambda_2$ at each point in $S_{\tau}$ (where $\lambda_i$ are the principal curvatures). This then shows that $|\Pi|^2=\frac{1}{2}H^2$ is constant on each $S_{\tau}$.

By equation (1.3) in \cite{HuiskenIlmanen}
\begin{equation}
\partial_{\tau} H = -\Delta_{S_\tau} (H^{-1})-(|\Pi|^2+\text{Ric}(\nu, \nu))H^{-1}
\end{equation}
which (since $H>0$ and constant on each $S_{\tau}$) means 
\begin{equation}
\partial_{\tau} H = (|\Pi|^2+\text{Ric}(\nu, \nu))H^{-1}
\end{equation}
which then (since $H$ and $|\Pi|$ only depend on $\tau$) implies $\text{Ric}(\nu, \nu)=constant$ on each $S_{\tau}$. Taking two traces of the Gauss equation and solving for Gaussian curvature $K$ of $S_{\tau}$ we obtain (c.f 5.5 in \cite{DisconziKhuri})
\begin{equation}
K=\frac{1}{2}R-\text{Ric}(\nu, \nu)+\frac{1}{2}H^2-\frac{1}{2}|\Pi|^2
\end{equation}
which shows that the Gaussian curvature is constant on each $S_\tau$. Therefore each $S_{\tau}$ is isometric to a round sphere with metric $r^2(\tau)d\sigma^2$ for the function $r(\tau)$ defined by the relation $4\pi r^2(\tau) = A(S_\tau) = Ae^{\tau}$. By noting the fact that $d\tau = 2r^{-1} dr$ and using the Gauss lemma, the metric can be written in the form
\begin{equation} \label{GaussLemma}
g=H^{-2}d\tau^2 +g|_{S_\tau} = \frac{4H^{-2}}{r^2}dr^2 + r^2d\sigma^2.
\end{equation}
Since $M_{CH}(S_\tau)=m$ for all $\tau$ we can solve for $H^2$ to find
\begin{equation}
H^2=\frac{4}{r^2}\left( 1- \frac{2m}{r} + \frac{q^2}{r^2}   \right)
\end{equation}
and combining with (\ref{GaussLemma}) we obtain the metric is Reissner-Nordstr\"om. Since it has a cylindrical end, it must be extreme, with $m=|q|$. 

A simple calculation then shows that
\begin{equation}
E= \frac{q_e}{r^2}\nu_r \quad \text{and} \quad B= \frac{q_b}{r^2}\nu_r
\end{equation}
where $\nu_r$ is the outward unit normal to the coordinate spheres, completing the proof. $\square$

\section{Proof of Theorem 2} \label{Proof of Theorem 2}

The idea is to use the same methods as those in the proof of Theorem 1. However, we no longer need to assume the electromagnetic fields are divergence free or that the generalized boundary is outer-minimizing. Begin with our initial data set $(M, g, E, B)$ and conformaly deform it by the same procedure as in section \ref{InitialConformalDeformation} to obtain an initial data set satisfying the strict charged dominant energy condition. We denote this deformed data set by $(M_{\varepsilon}, g_{\varepsilon}, E_{\varepsilon}, B_{\varepsilon})$ which has ADM mass $m_\varepsilon$ satisfying $|m_{\varepsilon}-m|\leq C\varepsilon$ and $|q_{\varepsilon}|=|q|$. 

Since we don't need the electro-magnetic fields to be divergence free, we don't need to apply any of the techniques found in section \ref{DivergenceConstraint}. Instead, we perform the gluing as in section \ref{Gluing} and restore the CDEC by solving the elliptic problem of section \ref{CDECConstraint} (in fact, the function $\zeta$ can now be chosen to be supported in $\Sigma(T)$). 

The set $\mathcal{F}$ is a minimal surface, and so we can take the outermost minimal surface with respect to the end $\hat{M}^+$. We then apply the ordinary positive mass theorem with charge \cite{GHH, KW}  to obtain
\begin{equation}
m_{\varepsilon, T} \geq |q|
\end{equation}
and taking the limit $T\rightarrow \infty$ we obtain
\begin{equation}
m_{\varepsilon}\geq |q|.
\end{equation}
Finally, taking the limit $\varepsilon \rightarrow 0$ we obtain
\begin{equation}
m\geq |q|.
\end{equation}
$\square$

\medskip

This gives an alternate proof to the same result which was first obtained in \cite{ChruscielBartnik} using spinorial methods. See also \cite{Abramovic}. We remark that our techniques should readily generalize to the non-time symmetric case.

\end{document}